# Energy saving mechanisms, collective behavior and the variation range hypothesis in biological systems: A review

Hugh Trenchard[1,*] and Matjaž Perc[2,3]

[1]805 647 Michigan Street, Victoria BC V8V 1S9
[2]Faculty of Natural Sciences and Mathematics, University of Maribor, Koroška cesta 160, SI-2000 Maribor, Slovenia
[3]CAMTP - Center for Applied Mathematics and Theoretical Physics, University of Maribor, Krekova 2, SI-2000 Maribor, Slovenia

*corresponding author: h.a.trenchard@gmail.com

Energy saving mechanisms are ubiquitous in nature. Aerodynamic and hydrodynamic drafting, vortice uplift, Bernoulli suction, thermoregulatory coupling, path following, physical hooks, synchronization, and cooperation are only some of the better-known examples. While drafting mechanisms also appear in non-biological systems such as sedimentation and particle vortices, the broad spectrum of these mechanisms appears more diversely in biological systems including bacteria, spermatozoa, various aquatic species, birds, land animals, semi-fluid dwellers like turtle hatchlings, as well as human systems. We present the thermodynamic framework for energy saving mechanisms, and we review evidence in favor of the variation range hypothesis. This hypothesis posits that, as an evolutionary process, the variation range between strongest and weakest group members converges on the equivalent energy saving quantity that is generated by the energy saving mechanism. We also review self-organized structures that emerge due to energy saving mechanisms, including convective processes that can be observed in many systems over both short and long time scales, as well as high collective output processes in which a form of collective position locking occurs.

**Keywords**: Energy saving mechanism, evolution, peloton, school, flock, herd, bioconvection

# 1. Introduction

For living organisms, the simple processes of living and moving to overcome gravitational and drag forces are energetically costly. Where possible organisms seek to reduce those costs, and in a purely physical sense, organisms can reduce those costs simply by staying sufficiently near others. In this way, nearby organisms couple their energy systems and facilitate energy saving mechanisms, and reduce their individual metabolic costs and/or travel faster in groups than in isolation. Examples of such mechanisms include air or hydrodynamic drafting, whereby drag forces are reduced in specific regions around massive bodies; vortice uplift whereby fluid motion effectively pushes organisms in their active trajectory; or the synchronization of body kinetics. For example, in human systems, drafting is a commonly observed energy saving mechanism among cyclists, generated when cyclists follow behind others in zones of reduced air-resistance. Power requirements when drafting a single rider are reduced by approximately 18 % at 32 km h$^{-1}$, 27 % at 40 km h$^{-1}$; and, when drafting in a group of eight riders, power requirements are reduced by as much as 39 % at 40 km h$^{-1}$ (McCole et al., 1990). See Alexander (2004) for an earlier review of drafting mechanisms.

Energy saving mechanisms have a fundamental role in evolutionary processes, and necessarily involve thermodynamic considerations. Thermodynamic approaches to evolutionary processes may be traced to Lotka (1922), who proposed that evolution and biological systems are mass- and energy-dependent, in flux and driven to increase. Schrödinger (1944) recognized that a highly organized biological system draws energy from its environment to generate within itself a lower entropy state, and suggested that the study of living systems must involve reconciling the self-organizing principles of biology with the laws of thermodynamics.

Prigogine and Lefever (1973) proposed that living systems are dissipative structures that maintain stable far-from-equilibrium states wherein the flow of energy and entropy fluctuate across system boundaries. Kaila and Annila (2008) developed equations to describe how evolution is a process of diminishing energy gradient and dissipation that shows up as overall decreases in mass. Similar processes include allometric scaling whereby body size correlates to energetic demands and energy dissipation is minimized (West et al., 1997; Brown et al., 1993). Further, thermodynamics is increasingly recognized as a framework for understanding ecosystem dynamics (Jorgensen and Svirezhev, 2004).

Apparently less studied or understood are the mechanisms that drive evolutionary systems toward their energy minima, and how these mechanisms propagate and affect the general process of evolution. Here we argue that energy saving mechanisms in biological systems have a fundamental role in evolutionary processes as they propagate and attract to lowest energy states, while conserving entropy. Further, we propose that energy saving mechanisms permit self-organizing structures at lower collective energy states and a broader range of heterogeneity among system organisms than might otherwise evolve without such energy saving mechanisms.

Our analysis begins with a general discussion of how energy saving mechanisms fit within far-from-equilibrium entropy physics. We evaluate the broad thermodynamic role of energy saving mechanisms in evolutionary processes. Then we identify a variety of energy saving mechanisms in coupled natural systems, and summarize research regarding these mechanisms. Further, we develop the variation range hypothesis, which Trenchard (2015) proposed as a general theory, having observed the close correspondence between the ~36 % range of the maximum power outputs of 14 cyclists in a mass-start race, the ~38 % energy saving implied by Olds' (1998) drafting coefficient, and the 39 % energy saving achieved by drafting at 40 km h$^{-1}$ within a group of 8 found by McCole et al. (1990).

In this review we are concerned with coupling dynamics that involve some energy saving for one or both members of the coupled system, generalized globally to multi-agent systems. We do not consider kinetic or morphological efficiencies that have evolved in individual animals, such as the streamlined body shapes of fish, or the wing apparatus' of birds.

## 2. Energy saving mechanisms and the second law of thermodynamics

Prigogine (1997) described far-from-equilibrium stable living systems and their entropy production as:

$$\Delta S_1 + \Delta S_e = 0, \text{ or } \Delta S_e = \Delta S_1 < 0 \tag{1}$$

where $\Delta S_1$ is the change of entropy within a given system, and $\Delta S_e$ is the change of entropy across system boundaries.

Further, in far-from-equilibrium dissipative structures, the distance from equilibrium is a critical parameter whereby probabilistic energy and entropy fluctuations of both internal and external origin cause the system to bifurcate within a critical regime (Prigogine, 1997).

Thus in any far-from-equilibrium system like evolution or other more narrowly-bounded biological systems, we might expect the presence of some entropy conservation mechanism, such as an energy saving mechanism. This mechanism allows the system to descend periodically to lower collective energy states, and reduces the energy required for the emergence of self-organized structures within the system. The entire system therefore retains a shorter distance from equilibrium than expressed in a higher energy system, and thus conserves entropy within the system, as illustrated in Fig. 1.

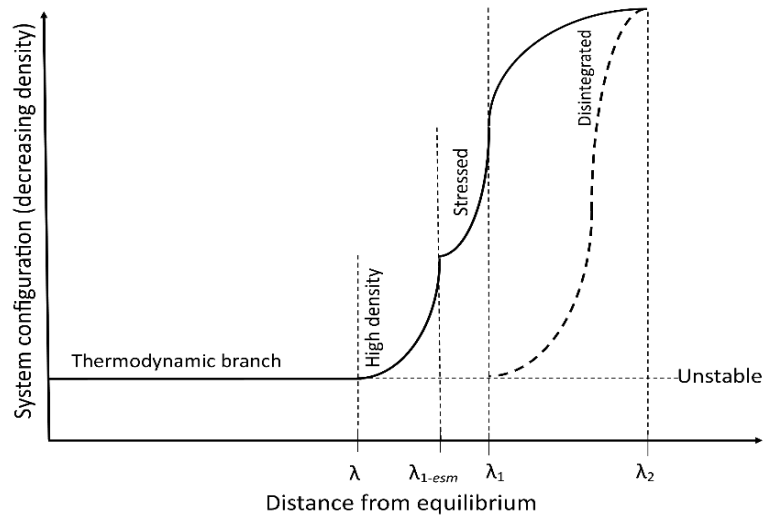

**Figure 1**. A far-from-equilibrium system with an energy saving mechanism. External energy crosses the boundaries of the system, driving it away from equilibrium and reducing entropy within the system. The system is susceptible to fluctuations from which different phases may emerge. Between $\lambda$ and $\lambda_1$ is the energy saving regime which permits phase changes at energy (temperature) gradients closer to equilibrium, thus conserving entropy within the system. Without the energy saving mechanism, the system transitions at higher energy gradients and farther from equilibrium, indicated by the dashed curve. Depending on the range of heterogeneity among system members, transitions beginning at $\lambda_1$ develop across a smaller energy gradient than within the energy saving region, and transitions from the high density phase to the disintegrated phase while bypassing the stressed phase, which is dependent on the energy saving mechanism. $\lambda_{1\text{-esm}}$, where the value 1-*esm* represents the percent energy saving generated by the energy saving mechanism, indicates a transition to a stressed phase. The energy saving mechanism permits a lower overall energy state for the system and maximizes entropy within the system. Figure inspired by analysis in Prigogine (1997).

The proposition that far-from equilibrium systems "fight" to conserve entropy was proposed by Schneider and Kay (1994), whose analysis of the second law of thermodynamics followed from a proof derived by Kestin (1966), whereby energetic systems,

> ...as they are moved away from equilibrium they will utilize all avenues available to counter the applied gradients. As the applied gradients increase, so does the system's ability to oppose further movement from equilibrium (Schneider and Kay, 1994).

Schneider and Kay (1994) identified Rayleigh-Bénard cell convection dynamics as an example of this entropic tension, a system that we shall explore further in this article. Rayleigh-Bénard convection cells self-organize in fluids at a critical temperature gradient, while additional increases in temperature gradient result in exponential heat dissipation; in other words, as the temperature rises, more work is required to increase the temperature gradient (Schneider and Kay, 1994). The authors proposed that the second law implies that self-organizing structures, such as the general process of evolution, will emerge as systems "attempt to resist and dissipate the external gradients that are moving them away from equilibrium" (p. 45).

Consistent with this proposition, we propose that biological systems are driven to exploit energy saving mechanisms as a means of resisting or narrowing system temperature gradients that would otherwise decrease entropy by increasing the energetic requirements of the system and dissipating entropy into the environment. In living systems, the temperature gradients are a function of the individual and collective metabolic processes of organisms within these systems. Individual temperature differentials change as environmental stressors drive organisms to increase or change their speed of movement (e.g. to run, swim, or fly faster), driving up or changing their metabolic outputs. Energy saving mechanisms therefore serve to reduce heterogeneous individual energy expenditures, which, without the presence of the energy saving mechanism, would be substantially higher.

Thus for many living systems, the energy required for the emergence of self-organized evolutionary structures like reproduction and speciation may in fact be *too high* without the presence of an energy saving mechanism; or, from an entropic standpoint, the system may never attain the necessary balance of energy inflow and entropy dissipation for self-organized evolutionary structures to emerge without an energy saving mechanism. The range of possible biological differences is therefore narrower without the energy saving mechanism, and such energy saving mechanisms may ultimately permit a greater diversity of life to evolve than without the energy saving mechanisms.

## 3. The variation range hypothesis

It becomes particularly evident that energy saving mechanisms are critical to the diversity of life if the maximal energetic capacities among coupled organisms are heterogeneous, spanning some relatively broad range of variation between individual maximal capacities. Obviously no species exhibits perfect homogeneity among its members, and we naturally expect some range of variation in physiological or metabolic capacities. The variation range hypothesis is the proposition that the range (percent difference) of heterogeneous metabolic variation within species tends to converge on the equivalent magnitude of the energy saving (percent difference) afforded by the energy saving mechanism. Makoto (1970) (as cited in Weihs (1973)) was an early proponent of this hypothesis in general terms, suggesting that hydrodynamic effects might determine a size range among fish schools of up to 50 %. Trenchard (2015) developed the hypothesis independently on the basis of simulated bicycle pelotons and their group sorting principles, which he demonstrated as self-organized effects that are independent of top-down anthropocentric tactical or strategic motivations.

This convergence in variation range occurs as organisms within a given system are driven by stressors to their individual, near-maximal physiological or metabolic outputs. At these near-maximum outputs, groups divide into sub-groups, the memberships of which are determined by the capacity of the weakest within each group to retain proximity to stronger members by exploiting the energy saving mechanism ($\lambda_1$ in Fig. 1). In this way the energy saving mechanism facilitates group cohesion; and obviously biological reproduction can occur only when group members stay within sufficient proximity to others. Without this cohesion mechanism, group divisions are likely to result in dispersion effects that are too great to permit reproduction and long-term, cross-generational survival. This of course does not exclude other mechanisms for group cohesion, but we suggest energy saving is a fundamental and primitive mechanism that is likely a precursor to other cohesion mechanisms that have evolved over geologic time. We refer to this as the **variation range hypothesis**.

The basic coupling equation describing group divergence and convergence and the variation range is

$$GCR = \frac{P_{pacesetter} * d}{MSO_{follow}}, \tag{2}$$

where *GCR* is "group convergence ratio"; $P_{pacesetter}$ is the power output, speed or metabolic output, of the pacesetter; *d* is the coefficient of the output of the follower, or beneficiary of energy saving, to the output of the pacesetter due to the energy saving mechanism; $MSO_{follow}$ is the maximal sustainable output of the follower, or beneficiary of the energy saving mechanism.

Here $(1 - d) * 100$ is the percent energy saving conferred by the energy saving mechanism, and the value that we propose is equivalent to the variation range of MSO among group members. When *GCR* > 1, coupled organisms diverge. Thus we see that among a collective of heterogeneous organisms with diverse MSOs, as long as their MSO is within the equivalent of (1-d) *100 of the leader's MSO, they will remain coupled as a cohesive unit, whereas any organisms whose MSOs > range equivalent of $(1 - d) * 100$, will decouple and form separate groups (*GCR* > 1), as shown in Fig. 2; and shown in Fig. 1 as the threshold between the stressed and the disintegrated regimes.

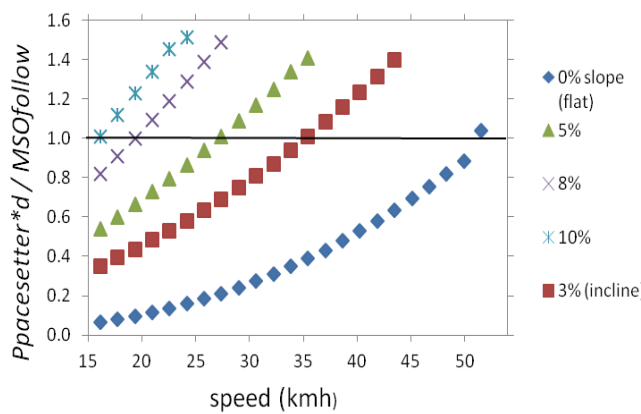

**Figure 2.** The group convergence ratio (*GCR*). The example shows the effect of the energy saving mechanism (drafting) on coupled cyclists. On a flat course, the equalizing effect of drafting allows a weaker cyclist with a maximum sustainable output (MSO) of 379W to maintain the pace of a stronger cyclist with MSO 525W up to ~50 km h$^{-1}$, when *GCR* = 1. On steeper terrain when equivalent power is required at speeds providing smaller drafting benefit, decoupling (*GCR* > 1) occurs sooner. On very steep hills when drafting is negligible, decoupling occurs immediately unless riders are nearly equal in strength. In any coupled system involving an energy saving mechanism, decoupling occurs similarly: a smaller energy saving mechanism means coupled organisms must be closer in strength to maintain coupling; greater energy saving means organisms with greater strength differential, within the variation range equivalent of (1-*d*) * 100 (given as percent), can stay coupled (reproduced with permission, from Trenchard (2013b)).

Since there are many reasons why organisms within living systems may be stressed to near maximal outputs, even if temporarily, it is apparent that Eq. 2 well describes a general group sorting mechanism, and that the energy saving in terms of percent, (1 − *d*) * 100, is a critical system attractor.

In this paper we will refer in some cases to speeds as indicative of $P_{pacesetter}$ and $MSO_{follow}$ values and energy saving percentages. Generally, power output, or metabolic parameters like maximum oxygen consumption, or body movement frequencies are preferable since speeds are not a good indicator of power requirements. This is because the power required to overcome fluid drag is proportional to the cube of the velocity, according to standard equations:

$$F_D = C_D \ \frac{1}{2} \ \rho \ AV^2 \ ,$$  (3)

where $F_D$ is the drag force, $C_D$ is the drag coefficient for the given fluid medium, $\rho$ is the fluid density, $A$ is the surface area facing direction of movement, and $V$ is the velocity of the object relative to the velocity flow of the medium. The power required to overcome the drag force is given by

$$P_D = F_D * V = \frac{1}{2} \ \rho V^3 \ AC_D.$$  (4)

However, in some cases, metabolic parameters are not available, and where necessary we will approximate $P_{pacesetter}$ and $MSO_{follow}$ and energy saving values using speeds independently of the other factors that determine power.

To summarize, we propose that this sorting mechanism, the bases of which are the energy saving parameter and the relative outputs of pacesetters and followers, is a fundamental evolutionary mechanism that determines specific boundaries for the diversity of life as groups divide and speciate over geologic time.

Our objectives now are to identify many of these energy saving mechanisms in nature, to examine the structures in which these mechanisms reveal themselves, and to identify the principles of and the evidence for the variation range hypothesis.

## 4. Energy saving mechanisms in non-biological systems

It is instructive to begin by identifying energy saving mechanisms in non-biological systems, the presence of which is evidence of their universal importance among dynamical systems generally. Indeed, given their presence in non-biological systems as a system attractor and generator of complex behavior, it is easy to speculate that they play a critical role in the primordial origins of life itself, although further analysis in this respect is beyond the scope of this paper.

Recheirt and Stark (2004) demonstrated that in a system of circling particles suspended in an optical vortice, two particles in close contact move faster than a single particle because the friction per particle is reduced, a form of drafting. Increasing the ring radius of the vortices increases particle alignment, which therefore move faster than in smaller radius vortices (Recheirt and Stark, 2004). Further, the collective motion of four particles in a chain result in self-organized positional exchanges between first and third particles (Recheirt and Stark, 2004). While the authors did not extend their study of positional exchange beyond four particles, we may extrapolate that larger aggregates will tend to exhibit self-organized positional exchanges involving more particles, and hence the emergence of dynamic collective patterns.

Recheirt and Stark (2004) did not report the percentage increase in speeds for the faster-travelling particles. In a similar study, however, Grujic and Helleso (2007) reported a pair of particles moving 15 % faster than a single particle; this increase in speed can be explained by hydrodynamic coupling of particles within the cohesive no-slip zone where the leading particle displaces fluid for the following particle and entrains its motion (Grujic and Helleso, 2007).

Šiler et al. (2012) studied polystyrene beads (520 nm diameter) in water. In Šiler et al.'s (2012) study, the beads were aligned in linear chains up to 18 beads long; particles in the centre of the chain caught up with leading particles and distanced themselves from tail particles. The authors reported increases in speed of up to 50% among faster moving chains.

Similarly, Wang and Guo (2015) reported the effects of "drafting, kissing, tumbling" (DKT) between two particles of equal size in a fluid medium. "Experiment A" featured two particles of equal size; "Experiment B" featured two particles of size ratio 2, with the smaller particle below the larger one; and "Experiment C" in which the smaller particle was located above the larger. Wang and Guo (2015) found that particles attract due to the wake and reduced drag generated by the leading particle. The authors found that speed of attraction was greatest for experiment B and slowest for experiment C; but there was a clear phenomenon involving the small particle accelerating to catch up and contact ("kiss") the leading particle. Wang and Guo (2015) reported that where particles were equal in size, the DKT process would repeat more than once, whereas the process occurred only once for particles of unequal size.

These dynamics indicate that not only does a leading particle in non-biological systems generate a low-drag wake in which a following particle may accelerate, but that ongoing coupled interactive processes can emerge from the energy saving mechanism. This supports the proposition that energy saving mechanisms are a fundamental and primitive process that precede other cohesive principles in evolutionary dynamics, and that dynamical processes emerge from these mechanisms.

For the purpose of demonstrating the ubiquity of energy saving mechanisms in coupled systems, we refer again to a non-biological system involving deformable bodies: flags. Zhu (2009) found that the relative drag forces of a leading and following flag in a downstream fluid flow depends on the Reynolds number (*Re*) of the viscous fluid. When *Re* is sufficiently high, Zhu (2009) found that the leading flag has less drag than the following flag, which Ristroph and Zhang (2008) defined as "inverted drafting"; at sufficiently low *Re*, the intuitively expected case of drafting is observed: the following flag has lower drag. Zhu (2009) found that the transitional value depends on the bending rigidity of the flags and the distance between flags. Ristroph and Zhang (2008) found that inverted drafting may be observed among flags which deform in response to the altered air flow caused by neighboring bodies, while rigid bodies like birds or fish do not.

## 5. Convective processes

Having referred to Schneider and Kay's (1994) application of Rayleigh-Bénard convection to entropic principles, now we examine convection patterns that emerge from collective interactions. This is distinct from convection occurring as a result of ambient heat or rising heat from body surfaces, which comprise a different field of study (Chappell et al., 1989). In Fig. 1, convective processes occur in the high-density regime.

Rayleigh-Bénard convection involves a rolling cycle of fluid particles vertically across a temperature gradient $\Delta T = T_2 - T_1$ which results in the rise of heated particles and the fall of cooling particles; the cyclical process involves gravity and buoyant forces and occurs at critical values of fluid density, temperature, and viscosity (Bergé and Dubois, 1984). In pans with large aspect ratio (cylinder diameter to height > 1), two dimensional cellular patterns emerge (Rivier, 1990). In cylindrical containers with aspect ratio approaching 1, Rayleigh-Bénard convection can generate a large-scale circulation loop with upflow and downflow appearing on opposite sides of the cylinder (Brown and Ahlers, 2009). Convection processes have also been demonstrated in granular flows which involve a central upward flow of vibrating grains, and their downward flow along container walls (Jaeger et al., 1996).

Although Rayleigh-Bernard and granular convection involve externally-applied energy gradients of heat or vibration, respectively, we propose that large-scale bioconvection emerges from internally generated temperature differentials between system organisms. Here we distinguish between large scale macroscopic bioconvection and small scale microscopic bioconvection.

The initial condition from which large scale bioconvection emerges is the inherent temperature differential between high and low energy positions that exist in the presence of an energy saving mechanism. In a system of coupled self-propelled agents, this temperature differential is unstable, and the broader convective process begins when organisms alternate positions across the temperature gradient, similar to the positional exchanges among four particles in an optical ring vortice (Reichert and Stark, 2004). In this way, dyadic, or locally coupled positional alternations, may be considered low-degree convection. In multi-agent collectives, large scale convection emerges as multiple "energetic" organisms move uni-directionally along lower density peripheral regions, and pass fatiguing organisms that move effectively backward within the system. This generates the internal rotational dynamics characteristic of convection. We propose that convection dynamics are a natural outcome of systems involving energy saving mechanisms and represent one phase or property of elementary evolutionary processes.

Mathematically, Rayleigh-Bernard convection generates vertically oscillated sinusoidal patterns, $\delta^{'}\cos\omega^{'}t$, described further by Rayleigh and Prandtl numbers and two dimensionless parametrically driven equations (Rogers et al., 2002). Such sinusoidal movement patterns have, for example, been observed in the collective movements of cyclists in pelotons, as shown in Fig. 3. Throughout this paper we will identify examples of low-degree (dyadic or locally coupled) convection and high degree convection (globally coupled) both in terms of large-scale macroscopic bioconvection and small-scale microscopic bioconvection.

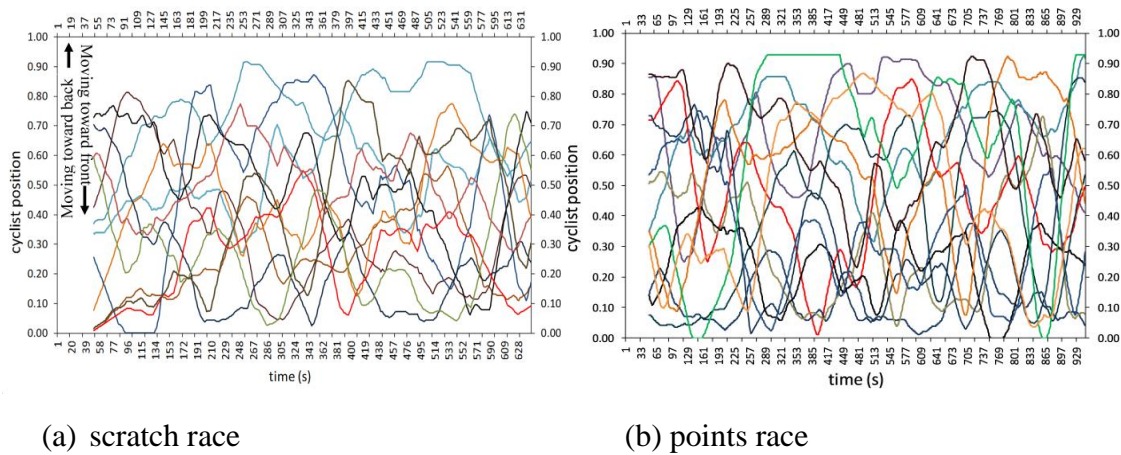

<div align="center">(a) scratch race       (b) points race</div>

**Figure 3.** Cyclists' positional profiles in two races. Moving averages (period 20) reveal approximately sinusoidal trajectories of cyclists in two directions moving from back to front, and moving from front to back. Rotations are approximately elliptical. In (a) 12 female cyclists in mass-start race on velodrome (circular track), fastest time wins; in (b) 14 (identical cyclists for scratch race +2), points accumulated every 5 laps by finishing order, most points wins (reproduced with permission, adapted from Trenchard et al., 2014).

Toner and Tu (1998) and Toner et al. (2005), in developing a fluid dynamical framework for flocks, referred to the "convective transport" of information and fluctuations in local velocities within flocks. However, the authors do not appear to have examined the specific rotational effects we identify here, and not in the context of an energy saving mechanism. Alberts (1978) referred to rotational movements of huddling rat pups as a convective current, but did not model the behavior. Bacteria have been show to exhibit bioconvection, as we discuss subsequently; but it appears that these processes have not been extrapolated to macroscopic biological systems.

Macroscopic biological systems may exhibit fluid-like properties akin to granular flows which do not require consideration of the surrounding medium if the grains are dry (Jaeger et al., 1996). In this way, macroscopic self-propelled agents (e.g. birds, fish, cyclists) may be modeled as the fluid particles that generate their own intrinsic viscosity and density independently of any fluid medium in which they thrive, as shown in Fig. 4. However, the factors determining bacterial bioconvection appear to be more complex (Wolgemuth, 2008).

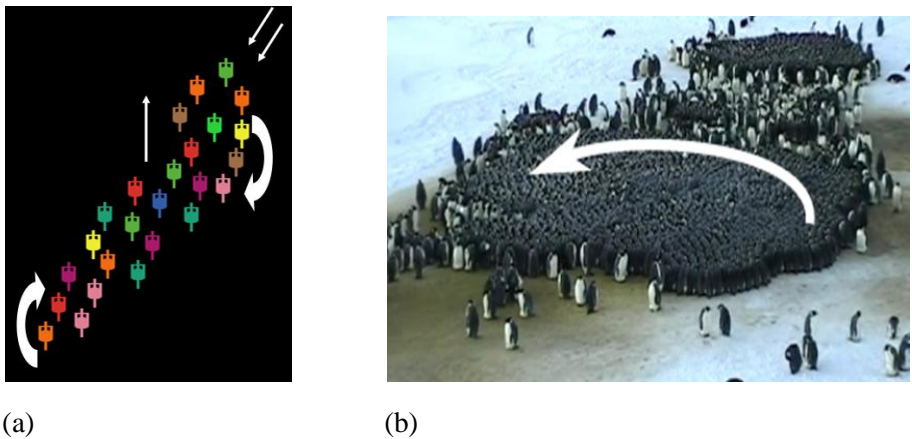

(a)                                    (b)

**Figure 4.** (a) Simulated echelon formation (simulation by H.Trenchard), using Netlogo (Wilensky, 1999). Cyclists proceeding north into a northeast wind begin a clockwise rotation; (b) Penguin rotations (reproduced with permission from A. Ancel, M. Beaulieu; and under *Creative Commons licence*, Gerum et al. (2013)). The cyclists and penguins are the particles of the fluid system, and the rotational dynamics are a form of convection.

## 6. Packing and phyllotaxy

We consider the role of Rayleigh-Bénard type convection as both an energetically minimal and optimal packing state in long timescale phyllotactic formations. We distinguish convection processes over long time scales from those of shorter or rapid timescales, such as cyclists, birds, or fish. These different processes may be considered two sub-categories of a broader class of energetic systems; and we may also consider that flocking systems and phyllotactic systems are fundamentally related by the presence of convective processes and energy saving mechanisms, differentiated largely in terms of timescale, as shown in Fig. 5.

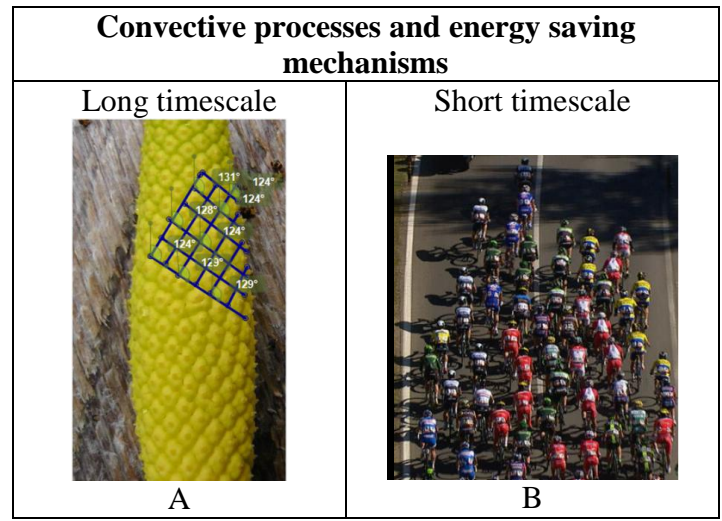

**Figure 5**. A broad class of convective systems with energy saving mechanisms. (a) Skunk cabbage spadix with two-dimensional Bravais lattice and approximate angles to the vertical, involving long timescale convective processes and inferred energy saving mechanism (angles may be distorted; image by Trenchard). (b) Peloton involving short timescale convective dynamics and energy saving mechanism of drafting. While avoiding collision and maintaining drafting positions, cyclists optimize packing formation in collective lowest energy state (reproduced with permission from race organizers Volta áo Algarve, Portugal 2014).

Hexagonal phyllotactic structures are typically observed in a variety of flora (Pennybacker, 2015). Lee and Levitov (1991; 1998) proposed that such structures represent the lowest energetic condition (the ground state) toward which mutually repulsive particles attract. Levitov (1991) further proposed that the phyllotactic ground state is the two-dimensional Bravais lattice (Bravais and Bravais, 1837) that matches the cylindrical optimal packing formation in which lines connecting central nodes form triangles. Nisoli et al. (2010) reproduced the energetic ground state experimentally using a magnetic "cactus" a structure of cylindrically-stacked magnets which self-organized to the collective lowest energy state in a Bravais lattice. Douady and Couder (1996) also demonstrated that ferromagnetic drops in a dome-shaped magnetic field form phyllotactic patterns, indicative of a minimal energy state.

Pennybacker and Newell (2013) observed that energetic minima and optimized packing in phyllotactic arrangements emerge from a dynamic process involving the physical motion and velocity of a moving annulus front which forms over the course of a few days – a much slower process than that seen in pelotons (e.g. Fig. 5a and Fig. 5b). Identifying sunflowers as an example, the authors noted that they form in two stages: first, flowers are initiated in an annulus, which moves farther out as the plant grows and configures in spiral patterns. At a certain point, central cells shift in phase such that they propagate florets or seed from the outside in, ring by ring. There is a defined velocity at the front of this inward propagating dynamic (Pennybacker and Newell, 2013).

These fronts comprise two kinds: pulled fronts and pushed fronts (Pennybacker and Newell, 2013). The first is determined by conditions ahead of the front; the second is determined by conditions behind the front and involves speeds greater than pulled fronts. For both fronts, the authors observed that the emerging pattern is a manifestation of the lowest possible energetic landscape. The authors thus described the floret generative process, a process of constant but comparatively slow motion, as one that inherently involves energetic minima. They suggested that hexagonal formations involve low local energy and high packing efficiency, a combination exhibited by the skunk cabbage spadix, as shown in Fig. 5a. Huddling penguins are also observed to pack in a hexagonal pattern (Zittebart, 2011); indeed, Waters et al. (2012) used the hexagonal pattern as a modeling parameter for penguin huddles, as shown in Fig. 12. Cyclists in pelotons pack similarly, albeit in faster continuous motion and at lower density than penguin huddles, and are positioned at angles to each other to avoid collision and to accommodate positional relocations ahead or behind neighboring riders, as shown in Fig. 4a and Fig. 5b.

In summary, the hexagonal floret pattern formation is an attractive low energy state, indicative of an energy saving mechanism that emerges over a lengthy temporal scale involving semi-permanent, quasi-static configurations. By contrast, systems like flocks or herds are temporally compressed (faster) versions of similar processes that, in terms of their relatively high speed movement, oscillate around their optimal collective configurations.

## 7. Bird flocks

We build our analysis by turning first to systems that are frequently observed in the ordinary course of human experience: birds and fishes. We then review the evidence for energy saving mechanisms in a number of other systems.

Lissaman and Shollenberger (1970) were perhaps the first to develop a robust quantitative power output model, based on a fixed wing, similar geometry analog of birds; although they did not say what kind of bird was analyzed. They reported a maximum possible energy saving from optimal flock positioning of about 70% for a flock of 25. They further observed that V formations permit an even distribution of drag. Birds in central positions within an in-line abreast formation (wing-tip to wing-tip) obtain greater energy saving than birds in peripheral positions, since central birds obtain the upwash advantage of two birds. In a group of 10, central birds obtain twice the energy saving of peripherally located birds.

By applying the aerodynamic model of Lissaman and Schollenberger (1970), Hainsworth (1987) reported an induced power (power to sustain sufficient lift to overcome gravity) reduction of 36 % for a total of 55 Canada geese (*Branta canadensis*) based on a median wing tip spacing of 19.8 cm – a critical parameter in determining vorticity since closer wingtip spacing means greater upwash effect and energy saving (Lissaman and Schollenberger, 1970). Lift forces that facilitate energy saving are also observed in locusts: specifically, a 16 % lift advantage occurs at up to 15 cm behind others (Camhi et al., 1995).

Cutts and Speakman (1994) reported that, based on measurements of wing-tip spacing among 54 skeins, Pink-footed geese (*Anser brachyrhynchus*) achieved mean power output reductions of 14 %; this is less than one-third of the maximum power output reduction, 51 % in a flock of 9 birds, a value calculated by Badgerow and Hainsworth (1981) by applying a modification of the Lissaman and Shollenberger (1970) model. The substantial inconsistency with the predicted values of Lissaman and Shollenberger (1970) was thought to be due largely to geese constantly adjusting positions resulting in relatively large variation in both wing-tip spacing, and corresponding flight time in non-optimal positions (Cutts and Speakman, 1994).

In addition to energy saving through formation flight, when flocks fly closer to smooth ground surfaces like water, they save energy through ground effect where wind speeds, wingtip vortices and downwash drag are reduced, allowing for increased lift and gliding times (Finn et al., 2012). For flocks of brown pelicans (*Pelcanus occidentalis*) exploiting ground effect, Hainsworth (1988) estimated energy saving between 49 % for gliding altitudes (mean ~33 cm), applying the aerodynamic model of Lissaman and Shollenberger (1970).

Although the energetic costs of birds can be predicted from aerodynamic models, such models predict only mechanical power requirements, which represent just 10-23% of total metabolic costs (Ward et al., 2002). This suggests more accurate models will be derived from empirical data that reflect aerobic costs, such as volume of oxygen consumption, or heart-rate. Ward et al. (2002) noted, however, that the correct relationship between heart rate and energy consumption is not established, although new methods to determine the energetic consumption of free-ranging animals are being developed (Nathan, 2012; Bairlein et al., 2015).

Weimerskirch et al. (2001) obtained heart rate data for a group of eight white pelicans (*Pelcanus onocrotalus*) which exhibited heart rate reductions that corresponded to 11.4-14.0% energy saving due to increased gliding time. Somewhat contrary to the findings of Hainsworth (1988), Weimerskirch et al. (2001) found no significant difference between heart rates of pelicans at 1.0 m above water versus those at 50 m above water. Hainsworth's (1988) findings were, however, in relation to altitudes closer to the water surface (0.2 to 1.0 m) where ground effect is greatest (Finn et al., 2012). It should be noted that Hainsworth's (1988) findings were based on Lissaman and Shollenberg's (1970) aerodynamic model and not actual data.

Using a computational fluid dynamical model, Maeng et al. (2013) found that Canada geese (*Branta canadensis*) can save 15 % energy by changing their wing morphology; and they can save 16 % by taking advantage of upwash vortices between 0 and 0.4 m wing tip separation. These values are in closer agreement to the values reported by Cutts and Speakman (1994) for Pink-footed geese (*Anser brachyrhynchus*) whose mean wing-tip spacings were observed to be 16.9 cm.

If we consider an approximate mean energy saving of 15 % for migratory birds, then according to the variation range hypothesis, if weaker fledglings are to sustain faster speeds of stronger adults, adults cannot fly at speeds greater than ~15 % faster than the younger birds' maximum sustainable speeds, averaged over migration duration. However, in this analysis it is important to consider the effects of higher speeds set by stronger birds which could force weaker birds to de-couple from the skein. This suggests that the *mean* energy saving of 15 % may not be the appropriate quantity by which to model flock cohesion, where cohesion is defined as birds that maintain sufficiently close proximity to obtain an energy saving benefit.

Conversely, the variability in energy saving (and, importantly, the maximum energy saving as a component of this variability) during flight may provide better information than the mean energy saving in understanding the nature of flock cohesion. Consider that at one instant, a weak bird may slip outside the optimal energy saving position and, if it is approaching its maximum output, it will be forced to decelerate relative to faster birds. However, if it adjusts to a near optimal position, it will resume equal pace to a stronger bird, albeit perhaps in a different relative position within the flock. In this way, the dimensions of the flock are important because they indicate how long a weaker bird can afford to decelerate before it must resume an optimal position, or risk being de-coupled from the flock. This can be shown according to the relation, as adapted from Trenchard et al. (2014):

$$T_{gap} = D_{last} / (V_p - V_{GCR > 1}), \tag{5}$$

where $T_{gap}$ is the time (s) for a bird that is decelerating relative to its flock mates to de-couple from the flock; i.e. when it drops beyond the last possible energy saving zone in the flock. We may refer to this as global de-coupling, versus local de-coupling when a bird suffers a temporary deceleration and falls outside the optimal energy saving zone of its nearest neighbors, but is sufficiently within flock boundaries to resume matching speeds with other nearby birds before it decouples from all possible energy saving zones within the group.

$D_{last}$ is the distance between the decelerating bird's position when it begins to decelerate, and the last energy saving zone in the flock, at which point it globally decouples; $V_p$ is the mean flock velocity determined by the pacesetter at the apex; $V_{gcr>1}$ is the velocity of the decelerating bird when it decouples locally, as shown in Fig. 6.

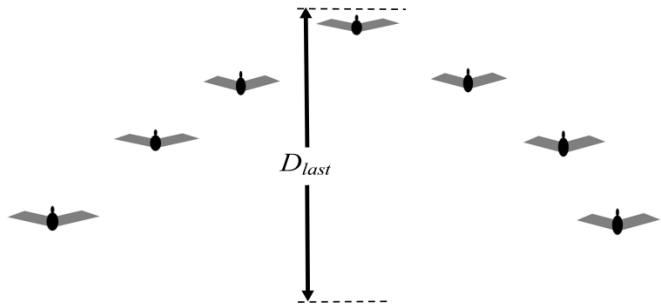

**Figure 6.** Illustrating $T_{gap}$. In this example, the skein travels vertically to the page. If the leading bird decelerates relative to the group, it has $T_{gap}$ (s) to find an energy saving position and resume the speed of the pacesetter, or it will decouple globally. While the bird will need to move at an angle to the vertical, $T_{gap}$ is still determined by $D_{last}$ as shown by the arrows.

It is clear therefore that the larger the flock, the more time a bird has to find an optimal position before it globally decouples. Similarly, the larger the flock, the more time there is for a skein to reduce its mean speed, which at sufficiently reduced speed will allow a locally decoupled bird to resume its coupled condition ($GCR < 1$). Eq. 6 and Eq. 7 and the discussion in the following section on fish schools, models this further.

Equation 5 implies that smaller flocks will tend to consist of birds within a narrower variation range, while larger ones will contain birds within a larger variation range. This permits predictions that: 1) when high speeds are sustained for $> T_{gap}$, groups will tend to sort into sub-groups whose variation range converges on the percentage energy saving benefit; 2) weaker birds will tend to be found in larger flocks whose variations in speed tend not to be of durations $\geq T_{gap}$.

This also predicts that flocks will tend to divide into sub-groups early in the migratory period when heart rates are high, after which flock size will stabilize as speeds or relative power requirements diminish. For example, Svalbard Barnacle Geese (*Branta leucopsis*) exhibit a relatively high heart rate at earlier stages of migration flights (mean 317 beats min$^{-1}$) compared to the landing stage (mean 226 beats min$^{-1}$) (Butler et al. 1998), suggesting early stage group sorting. The reduction in heart-rates over the migratory period is likely caused by the fact that stronger birds will have fatigued after setting a faster pace earlier in the migration, while fledglings and weaker birds will have saved energy from remaining in energy savings positions; thus, overall, there has been an equalization of maximal sustainable power outputs among group members.

Among formations in which some positions offer greater benefits than others, we might expect some rotation between higher cost and lower cost positions; or, we might expect weaker birds to spend longer periods in more advantageous positions, such as couched within the V formation where birds effectively assume an in-line, abreast formation. In fact Lissaman and Shollenberger (1970) suggested this possibility, noting that young or weak birds may tend to spend more time in positions of greatest energy saving. Bill and Herrnkind (1976) also recognized this possibility in the context of the queue formations of spiny lobsters (as we will discuss subsequently) as did Makoto (1970) in the context of fishes.

Consistent with the Lissaman and Shollenberger (1970) proposition, Voelkl et al. (2015) reported that Northern bald ibis' (*Geronticus eremita*) swap in-wake positions (dyadic reciprocation) which permit lower metabolic output, a situation described by the authors as a cooperative mechanism. Voelkl et al. (2015) argued such positional reciprocation is especially beneficial for juvenile birds which experience high mortality rates. We also observe that these dyadic positional rotations are consistent with a convection process, in which the shift from a higher metabolic position to a lower one is a small-scale convective roll from a higher temperature position to a lower one.

In terms of juvenile mortality rates, Menu et al. (2005) reported a pair of pertinent observations: snow geese (*Chen caerulescens atlantica*) goslings fledge (develop flying feathers) when their body mass rises to 76% of its adult maximum; and migratory departure occurs within a few days of fledging (citing Blouin,1996). Menu et al (2005) noted that fledglings "too weak or too light," (presumably lighter than 76 % body mass of adults), will not survive. To the extent that body mass correlates with strength, the authors did not explain why fledglings that are approximately 24 % weaker than adults can in fact survive, since the expectation is that without some mechanism to permit speed equalization, weaker fledgling would simply be dropped from the group if pacesetter speeds exceed by 24 % the maximum sustainable fledgling speed.

Even when adults relax their own pace by 24 % to accommodate fledglings, the absence of an energy saving mechanism would necessitate fledglings traveling at their maximum pace while adults are flying comfortably. We would expect fledglings to fatigue faster than adults in such a case, resulting in even greater differentials in maximal sustainable outputs between adults and fledglings. The fact that this differential does not arise, and that weaker birds frequently arrive at destination points along with stronger ones, can be explained by the existence of an energy saving mechanism. Thus on application of these values, the variation range hypothesis predicts that an energy saving of 24% is achieved by the aerodynamic advantages of flock positioning – advantages that are likely exploited largely by fledglings.

This 24 % difference between adult and fledgling metabolic capacity obviously exceeds the approximate proportionate mean energy saving value of 15 % identified above. The 24 % quantity may be closer to the maximum energy saving quantity; this would more accurately reflect the decoupling threshold and a corresponding size variation among groups that form as a consequence of decoupling, than does the mean energy saving of 15 %.

Survival rates among young geese are nonetheless determined by a combination of factors, including body mass, fledging date, and air temperature (Menu, 2005). Therefore, a more complicated analysis of the effects of fatigue rates and energy consumption factors on geese' current maximal sustainable outputs is required to accurately evaluate the relationship between goose size and/or strength variation ranges and the energy saving quantity, and to establish their precise proportions. Regardless, the absence of this precision does not weaken the logic of the variation range hypothesis and the basic relationship it establishes between maximal sustainable outputs modulated by the energy saving benefit, and the variation range of outputs within a given group.

It is important to state that not all flocking behavior involves energy saving. The energy costs of turning motions are high (Wilson et al., 2013; Amélineau et al., 2014); indeed pigeon flocks, which exhibit frequent turning behavior, have been shown to incur higher energetic costs due to the increased g-forces of banking (Usherwood, 2011). Also, Baldaccini et al. (2000) surmised that flocks of wild rock doves *(Columba livia)* engaging in direct "commuting" flights up to 18.9 km, flying in three-dimensional clusters (as opposed to flat V-formations) of up to 40 or more birds, did not entail an aerodynamic advantage because of the lack of precise spacing between individuals. However, we suggest that in a rotating three-dimensional cluster, birds are likely to pass frequently in and out of energy saving upwash vortices, while avoiding high turbulence positions that are directly behind others, thereby avoiding collisions and aligning trajectories for continuous passing motion. We have not seen studies that sum the costs/benefits of passing in and out of upwash positions in clustering flocks; hence, when a cluster flock flies in point-to-point trajectories without a high degree of banking behavior, the aerodynamic energy saving advantages of cluster flight remain unknown.

Flack et al. (2013) have, however, proposed that flock clusters provide navigational efficiencies, which is yet another kind of energy saving mechanism that leads to reduced flight distances. This is similar to the leader/follower energy saving dynamic due to attentional efficiencies, as demonstrated by Piyapong et al. (2007). This kind of navigational efficiency is also similar to Wilenksi's (1997) ant-following algorithm in which following ants wait a short time before advancing across the secant of the path taken by leaders, a strategy which eventually leads to the shortest distance between travel points.

## 8. Fish schools

Hydrodynamic drafting and Karman gait (kinematic synchronization with vortices) (Liao, 2007), are now well-established as energy saving mechanisms among schooling fishes, although the extent of their effect on schooling behavior has been a matter of controversy (Pitcher and Parrish, 1993; Lopez et al., 2012). Breder (1965) was perhaps the first to identify energy saving mechanisms in fish schools. Weihs (1973) predicted energy saving of 40-50 % for fish in optimal positions, while earlier Zuyev and Belyayev (1970) reported tail beat frequency (TBF) reductions of 15-29 % for trailing horse mackerel (*Tracurus mediterraneus*) over leading fishes.

More recently, Herskin and Steffensen (1998) reported that sea bass (*Dicentrarchus labrax*) in trailing positions during natural swimming activity showed 9-14 % reductions in TBF, and a 9-23 % reduction in oxygen consumption. Marras et al. (2015) reported that Grey mullets (*Liza aurata*) swimming within sufficiently close proximity to others generated reductions in TBF up to 28.5 % at 10 cm.s$^{-1}$, relative to TBF when swimming in isolation. Rates of oxygen consumption for fishes benefitting from hydrodynamic drafting were lower by between 8.8 % and 19.4 % relative to swimming alone (Marras et al, 2015).

Svendsen et al. (2003) studied groups of eight roach (*Rutilus rutilus*) with body-length variations between 7 and 14 %, (which is within the body-length variation of natural schools), and mass variations between 17 and 41 %. The authors increased flume current speeds from $1_{LTMS}$s$^{-1}$ (mean body-length per second) to $4_{LTMS}$s$^{-1}$, and observed that between 2 and $4_{LTMS}$s$^{-1}$ fish positioning within the group of eight stabilized while trailing fish obtained TBF reductions of 7.3, 11.9, and 11.6 % for 2, 3 and $4_{LTMS}$s$^{-1}$ respectively.

Pitcher and Parrish (1993) were critical of arguments for hydrodynamic advantages generally, arguing that the evidence did not bear out some of Weihs' (1973, 1975) pioneering wake theory predictions, including: precise distances between fish, the presence of tail-beat anti-phase, and the existence of the "scramble for good positions", which Pitcher and Parrish (1993) considered to be evolutionarily unstable. Pitcher and Parrish (1993) further considered that any alternations between costly and optimal positions required altruistic behavior; a behavior which the investigators implied was unlikely. Their criticism, however, ignores the self-organized nature of fish positioning which naturally emerges, without altruism, from metabolic stresses that necessitate adjustments in speeds according to drag and other physical forces.

Liao (2007) cautioned that a reduction in TBF could arise from a combination of wall effects (Webb, 1993), flow refuging, and vortex effects, and not from vortex effects alone. Further, the hydrodynamical benefits of three-dimensional positioning have not been well-studied (Liao, 2007). Weihs (1973), however, suggested that in addition to horizontal energy saving, vertical stacking and vortice upwash from fin tips may provide energy saving. Blake (1983) noted that downwash forces may neutralize any vertical energy saving mechanisms.

Nevertheless, in a subsequent study Marras et al. (2015) found that, except for an approximate 30° cone directly behind others where there is increased turbulence, some energy saving was present in all planar regions around individual fishes, even directly ahead of others, as shown in Fig. 7. Using computational fluid dynamics simulations, Helmelrijk et al. (2015) found that fishes can save energy even directly behind others if followers bend their heads sideways and capture energy from the vortex flows.  In similar fluid dynamical simulations, Becker et al. (2015) found energy saving in groups of flapping fishes would enhance speed and reduce power output.

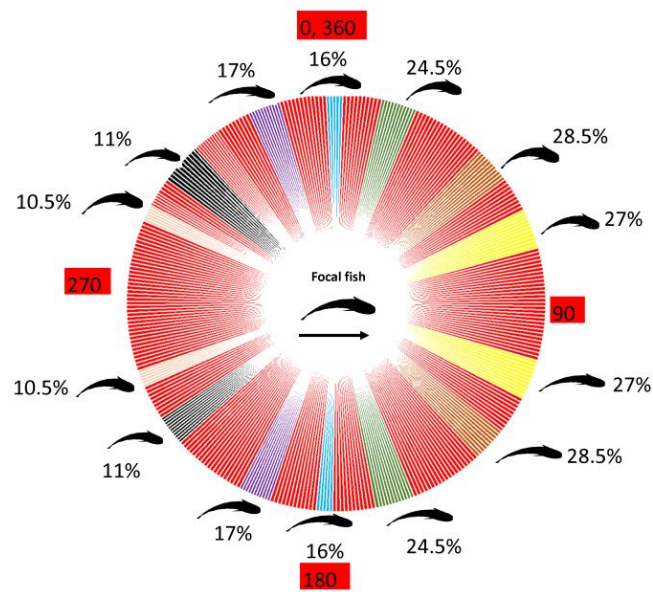

**Figure 7.** Overhead view showing energy saving zones in terms of tail-beat frequency (values from Marras et al. (2015)). Energy saving coefficients within red (color online) zones are the averages of the zone coefficients on either side, except for the zone in the 30° region on either side of 90° where there may be negative energy saving because of turbulence. Using computer simulations of fluid dynamics, Hemelrijk et al. (2015) found that even in this zone, if fish bend their heads sideways, they may capture energy from vortices.

The Marras et al. (2015) findings expand earlier findings by Killen et al. (2011), who reported that trailing fishes within a school required fewer tail beats to swim at the same speed as fishes at the front of schools. Due to energy saving in the entire circumference around nearby fishes, the authors proposed that such reduction in locomotion costs is one of the primary drivers of fish schooling behaviors.

### 8.1. Convective and protocooperative dynamics

We have proposed that an energy saving mechanism is likely to generate a convective phase for organisms when collective outputs enter a critical range. Among schools, convective behavior is likely observed when fishes alternate between high cost and less costly positions. This behavior also describes a form of cooperation.

In their phase transition model, Becco et al. (2006) observed that the cross-sectional distance between fishes at school boundaries (schooling width) describes the degree of fish cooperativeness, but they did not state what factors define cooperativeness. Their model does permit analysis of school structure based on relative fish velocities and, by extension, durations spent by fishes in particular positions within schools. This kind of analysis may assist in more precisely defining cooperativeness and 'protocooperative behavior' (Trenchard, 2015) within schools because it will show the extent to which fish share the most costly positions – or, those positions having higher drag; or greater vulnerability to predators (although this factor is not considered under protocooperative theory).

Protocooperative behavior is defined by two phases of behavior, a comparatively low collective output phase in which biological agents within a system are capable of sharing highest-cost front positions, and a high collective output phase in which agents can sustain the speeds of faster agents, but cannot pass them in order to share the highest-cost front positions; the threshold between these phases is the 'protocooperative threshold' (Trenchard, 2015) ($\lambda_{1-esm}$ in Fig. 1).

Along these lines, Domenici et al. (2002) quantified the nature of herring (*Clupea harengus*) positional shuffling behavior and O-turns, while controlling the water oxygen concentration necessary to induce fish hypoxia. O-turns occur when fish in front or near-front positions turn 180°, proceed in the opposite direction, then turn 180° again at the rear of the school; shuffling involves adjustments in relative positions uni-directionally (Domenici et al. 2002). Svendsen (2003) observed similar O-turn behavior among roach (*R. rutilus*).

Pitcher and Parrish (1993) might argue that altruism is required to explain the presence of O-turn and shuffling behaviors. On the contrary, however, O-turn and shuffling behaviors are evidence of protocooperative behavior in which high cost positions are shared when collective outputs are sufficiently low (Trenchard, 2015). These behaviors are also evidence of convective behavior similar to Northern bald ibis dyadic rotations (Voelkl et al., 2015), Emperor penguin huddle rotations (Le Maho, 1977), and spiny lobster (Bill and Herrnkind, 1976) lead position rotations.

We suggest O-turns and shuffling represent low degree convective behavior because they appear to involve power output (analogous to temperature fluctuations) in which fishes shift between high drag front positions and lower drag, following positions. Similarly, these positional alternations are a form of protocooperative behavior in which we may predict that fishes, at lower outputs, are capable of sharing high-cost positions (the high-density regime in Fig. 1, or the convective phase), but at a critical speed or output threshold are unable to engage further in sharing these positions and will maintain relative positions by drafting (Trenchard, 2015) (the stressed regime in Fig.1, or the position locking phase).

Domenici et al. (2002) found that in conditions of higher oxygen saturation, the herring engaged in O-turns more frequently; however, as hypoxia increased, they reduced O-turn frequency. Thus at oxygen saturation below 25 %, O-turn frequency occurred at less than 1/3 of reposition manoeuvres, compared to more than ½ in conditions of normal oxygen saturation (Domenici et al., 2002). The authors suggested this may be due to due to a pair of factors: the high cost of relatively swift O-turn manoeuvres, and the fact that, in larger schools, rear positions are further reduced in available oxygen.

We propose an alternative hypothesis, consistent with protocooperative behavior (Trenchard, 2015): the threshold at which O-turn frequency drops is a function of individual fish hypoxia relative to maximal individual oxygen uptake capacity, modulated by the output required for the position whether it is a high cost front position, or a lower cost drafting position. This alternative hypothesis is supported theoretically by Killen et al.'s (2011) results which demonstrated that grey mullet (*L. aurata*) with the highest aerobic and metabolic capacities tended to shift toward the front as current speeds were increased, while the converse was true for weaker fish.

Increase in hypoxia in rear positions can be discounted for the moment because this effect was unlikely in the small schools of 19-22 fishes studied by Domenici et al. (2002). Thus, fishes in optimal energy saving positions will experience a reduced incidence of hypoxia because oxygen uptake requirements are lower. The front positions therefore represent the highest energy costs due to highest combined drag and hypoxia. At sufficiently high oxygen saturation, fishes of all aerobic capacities may trade-off between the highest cost front positions; as oxygen saturation decreases, however, only the fishes with the greatest aerobic capacities will be capable of sustaining the highest cost front positions, while weaker fish can maintain their positions in lower cost energy saving positions, a determination which is consistent with the observations of Killen et al. (2011). When an unstable threshold condition is reached, if the strongest fishes choose to O-turn, then the entire formation collapses; and, indeed, Domenici et al., (2002) appear to have described such a breakdown in formation in the presence of very low oxygen saturation levels, between 12 and 25 %, a range at which fishes began swimming in different directions.

Domenici et al. (2002) found that fishes tended to shuffle at relatively constant rates over different oxygen saturation levels. Svendsen et al. (2003) reported continuous positional movement over three speeds (lowest, intermediate, and highest), with greatest positional movement rates at the intermediate speed, and somewhat lower movement at the lowest and highest speeds. Domenici et al.'s (2002) findings are somewhat inconsistent with protocooperative theory (Trenchard, 2015), which suggests that the rate of positional change falls at or near the protocooperative threshold, which occurs at a comparatively high speed when the metabolic capacity or power at which organisms can sustain pacesetters' speeds by drafting, but cannot pass them (the stressed regime in Fig.1).

On the other hand, Svendsen et al.'s (2003) findings are not inconsistent with protocooperative theory since we would expect higher positional movement at the intermediate speeds (the cooperative or convective phase; the high density regime in Fig.1), and reduced movement at higher speeds (the position locking phase; the stressed regime in Fig.1). Furthermore, the position locking phase is comparatively unstable and determined in part by fatigue rates of pacesetters and, at sufficiently high speeds, even the strongest fish are forced to decelerate. This may result in the strongest fishes reducing speed to shift their positions backward to lower cost positions, which potentially leaves weaker fishes exposed to the highest cost positions. In this state, the overall effect is collective deceleration and a breakdown of the higher speed formation, similar to the noted collapse of the fishes' formation between 12 and 25 % oxygen saturation in the Domenici et al. (2002) study.

## 8.2.   Evidence for the variation range hypothesis

It is well known that fish often school with others of similar size (Hoare, et al., 2007). Moreover, fish speeds tend to increase in correspondence with fish size (Krause et al. 2005); hence, a natural explanation for fish school sorting is that faster, larger, fish tend to aggregate together and, likewise, so will slower, smaller, fish. Pitcher and Parrish (1993) were critical of this hypothesis because, as they apparently reasoned, groups are more likely to divide and form sub-groups according to speed grouping (observed where there are very large size and age differences among fish), as opposed to size sorting within larger cohesive groups (i.e. sorting within groups without dividing). Foraging competition, individual decisions to swim next to neighbors of similar size (Pitcher and Parrish, 1993) and chemical cues (Ward and Currie, 2013) are among the explanations for school sorting by size.

Pitcher and Parrish (1993) did observe that among schools there may be body-length differences of up to 30 % within the same shoal or school. According to the variation range hypothesis, to the extent that this body-length or overall size variation correlates to the maximal metabolic capacities or speeds (Weihs, 1973; Krause et al., 2005) of individual fishes, this 30 % variation in body-length leads to the prediction that the energy saving quantity is ~ 30%.

Conversely, strictly based on the energy saving quantity, the variation range hypothesis predicts that for the Grey mullet (*L. aurata*) schools studied by Marras et al. (2015) which exhibit an approximate mean 20 % reduction in oxygen consumption, the variation in individual maximal metabolic capacities among school members is approximately 20 %. This value is remarkably near the 30 % body-length range indicated, given the various possible imprecise parameter measurements we discuss in this article.  These values support the proposed school sorting processes, particularly during migration periods when stronger fishes are expected periodically to drive speeds sufficiently high to cause group sorting; these values also suggest that body-length converges to this 20% range over geologic timescales. If groups, sufficiently large for viable reproduction, separate permanently as a result of these sorting processes, speciation is a natural consequence.

The work of Koutrakis et al. (1994) included extensive measurements of the body-lengths of different mullet species. The authors collected specimens intermittently in two-day periods over two years.  Given that samples were taken over the course of two days and therefore not all sample sets were composed of fish from the same schools, the reliability of the data as evidence for the variation range hypothesis is limited. Also, school sizes from which samples were collected were not given.  Nonetheless, such data does reflect long term adaptation, and to that end we have excerpted size data from the Koutrakis et al. (1994) study and included percent variation ranges, as shown in Table 1 below.

| months | Liza aurata (157) | | | Liza ramada (550) | | | Chelon labrosus (745) | | | Liza saliens (641) | | | Mugil cephalus (24) | | |
|---|---|---|---|---|---|---|---|---|---|---|---|---|---|---|---|
| | min | max | range % | min | max | range % | min | max | range % | min | max | Range % | min | max | range % |
| 4/89 | 32.0 | 46.0 | 30.4 | 16.7 | 29.3 | 43.0 | 13.6 | 22.3 | 39.0 | 21.8 | 126.0 | 82.7 | 32.0 | 46.0 | 30.4 |
| 5/89 | | | | 32.7 | 43.9 | 25.5 | 14.8 | 38.3 | 61.4 | 99.0 | 99.0 | 0.0 | | | |
| 6/89 | | | | 53.0 | 75.0 | 29.3 | 19.7 | 58.0 | 66.0 | 102.0 | 124.0 | 17.7 | | | |
| 7/89 | | | | | | | 38.5 | 64.7 | 40.5 | 30.7 | 30.7 | 0.0 | | | |
| 8/89 | | | | | | | 58.0 | 83.0 | 30.1 | 15.6 | 59.5 | 73.8 | | | |
| 9/89 | | | | | | | 66.0 | 107.0 | 38.3 | 17.2 | 53.0 | 67.5 | | | |
| 10/89 | | | | | | | 74.0 | 117.0 | 36.8 | 19.2 | 56.8 | 66.2 | 20.4 | 20.4 | 0.0 |
| 11/89 | 16.9 | 24.3 | 30.5 | | | | | | | 19.0 | 45.5 | 58.2 | 19.9 | 22.9 | 13.1 |
| 12/89 | | | | | | | | | | 18.3 | 54.5 | 66.4 | 20.3 | 25.5 | 20.4 |
| 1/90 | | | | 14.6 | 18.3 | 20.2 | | | | 17.7 | 24.2 | 26.9 | | | |
| 2/90 | | | | | | | | | | | | | | | |
| 3/90 | 22.5 | 35.5 | 36.6 | | | | | | | 41.8 | 41.8 | 0.0 | | | |
| 4/90 | 28.0 | 35.5 | 21.1 | 24.1 | 29.7 | 18.9 | 19.2 | 22.7 | 15.4 | 24.6 | 53.6 | 54.1 | | | |
| 5/90 | | | | 28.5 | 44.4 | 35.8 | 13.0 | 29.0 | 55.2 | 29.4 | 33.1 | 11.2 | | | |
| 6/90 | | | | 49.0 | 60.8 | 19.4 | 25.1 | 52.0 | 51.7 | 67.5 | 99.0 | 31.8 | | | |
| 7/90 | | | | | | | 43.8 | 62.0 | 29.4 | 11.6 | 99.0 | 88.3 | | | |
| 8/90 | | | | 107.0 | 108.0 | 0.9 | 79.0 | 95.8 | 17.5 | 9.3 | 102.0 | 90.9 | | | |
| mean | | | 29.7 | | | 24.1 | | | 40.1 | | | 46.0 | | | 16.0 |

**Table 1**. Body-length ranges for five species of mullet. Table modified from Koutrakis et al. (1994) (any missing data also missing in the original).

We also note that an exact correspondence between the variation range and the energy saving quantity is unlikely due to imprecise measurements of the energy saving quantity; imprecision in this quantity arises from both the lack of correlation of body-length and strength and the fact very large schools are likely to increase the variation range body-length (Eq. 7). However, the basic mechanism of size sorting occurs when speeds are driven high enough to separate weak fishes from the group. The actual reasons for high schooling speeds, whether because of migration, predator avoidance, hunger, or something else, are irrelevant to the production of sorting as a consequence of school speeds that are driven to a decoupling threshold.

In a study of the oddity effect, a phenomenon in which different species will intermingle when their body-lengths are similar, Ward and Krause (2001) reported fish body-lengths among 13 complete shoals composed of between 5 and 78 European minnows (*Phoxinus phoxinus*). From the authors' data and statistics, we estimated body-length ranges as percent differences using the range rule: range = s.d. * 4; max = mean + (range/2), as shown in Table 2.

| shoal | shoal number | mean body-length (mm) | s.d. | range | range % |
|---|---|---|---|---|---|
| 1 | 10 | 44.35 | 3.38 | 13.52 | 26.45 |
| 2 | 7 | 40.71 | 3.9 | 15.6 | 32.16 |
| 3 | 67 | 12.81 | 1.26 | 5.04 | 32.88 |
| 4 | 8 | 46.25 | 5.5 | 22 | 38.43 |
| 5 | 10 | 44.3 | 7.04 | 28.16 | 48.24 |
| 6 | 5 | 42 | 2.92 | 11.68 | 24.41 |
| 7 | 22 | 51.36 | 5.8 | 23.2 | 36.85 |
| 8 | 8 | 50.25 | 4.71 | 18.84 | 31.57 |
| 9 | 54 | 18.5 | 3.85 | 15.4 | 58.78 |
| 10 | 78 | 48.8 | 4.92 | 19.68 | 33.56 |
| 11 | 72 | 15.81 | 3.35 | 13.4 | 59.53 |
| 12 | 10 | 53.5 | 6.92 | 27.68 | 41.10 |
| 13 | 14 | 17.29 | 2.89 | 11.56 | 50.11 |
| **mean** | | | | | **39.54** |

**Table 2**. European minnow (*P. phoxinus*) body-length of ranges (percent difference), estimated from Table 1 in Ward and Krause (2001). Mean range = 39.54 %.

DeBlois and Rose (1996) reported the sorting of very large shoals into sub-groups by size. For a shoal of migrating Atlantic cod (*G. morhua*) that spanned 22 km and tracked for 36 days, the authors reported the presence of sub-groups for which the mean body-length of fishes in the leading group was 47.2 cm, and the mean body-length of fishes in the hindmost group was 43.6 cm, as shown in Fig. 8. DeBlois and Rose (1996) did not report maximums and minimums, but using their data and statistics, we estimated length ranges of 36-53 % for the five categories reported.

The sorting reported by Deblois and Rose (1996) supports the predictions of the variation range hypothesis.  In Fig. 8, we summarize the observations of Deblois and Rose (1996), and show its similarity in structure to the peloton simulation results of Trenchard et al. (2015), in which groups of 100 simulated cyclists with a 36 % variation range of maximal sustainable power outputs (MSO) were driven to near maximal outputs for periods that corresponded to actual race conditions. By the end of each simulated test trial, the original groups had divided and sorted according to relative strength, with new groups containing riders whose MSOs spanned a narrower range.

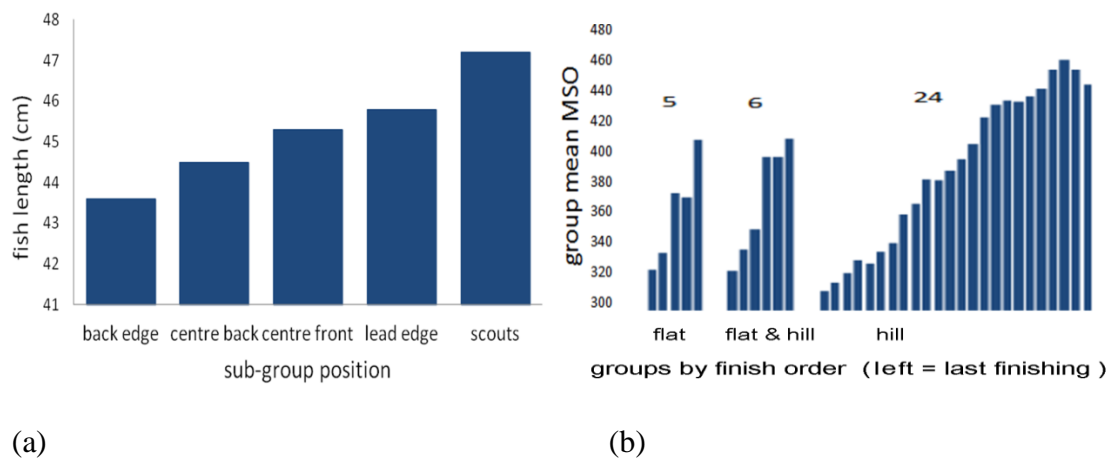

(a)                                                    (b)

**Figure 8**. (a) Sub-group positioning of large shoal of Atlantic cod, from data in DeBlois and Rose (1996). (b) Peloton sorting. Simulated cyclists were randomly assigned maximum sustainable power outputs (MSO) between 305W and 479W (range 36 %), power outputs converted from cyclists 200 m times. For each trial, groups of initial size 100 were run for simulation time ~ 16 minutes. Trial speeds were based on speed profile from an actual mass-start flat race, varying for (1) a flat course (2) the presence of a single hill (4 %) for 19 s and otherwise flat, and (3) the entire trial on a hill (3 %). At trial completion, for a flat course and flat & hill, the original group of 100 divided into a small number of groups, with weaker groups finishing last and each group with narrow MSO range  (i.e. < 36 %).  Due to the continuous high output required on the 3 % hill and reduced drafting benefit, the original group of 100 sorted into 24 groups (c.f. Fig. 2) (reproduced with permission, from Trenchard et al., 2015).

In contrast to this mechanism by which fishes sort according to size and/or strength, Ward and Currie (2013) proposed that fishes are able to locate others of similar size through a chemical cue mechanism. We suggest that although chemical cues may have evolved to become a component of school sorting and are not mutually exclusive of sorting by strength, the sorting principle proposed by Trenchard et al. (2015) in the context of pelotons, and as a consequence of the variation range hypothesis, is a simpler and more intuitive explanation. Since we expect that individual fish size will correspond to metabolic capacity, we argue that when schools are driven to speeds corresponding to the maximal output of weaker fish, even for relatively short periods, the schools will sort into sub-groups defined by a specific, narrow range of maximal outputs and size, such as shown in Fig. 8. As the variation hypothesis range predicts, this narrow range will approximately correspond to the energy saving quantity, modulated by the school size.

The variation ranges inferred from Koutrakis (1994), Ward and Krause (2001), and DeBlois and Rose (1996) all indicate variation ranges higher than predicted by a one-to-one correspondence between the species-specific energy saving quantities and the variation range. We should not, however, generalize the energy saving quantities of certain species to those of others. For species in which energy saving quantities co-exist with body-length variation range data, the correspondence between these parameters appears closer, as shown in Table 3.

| | Energy saving TBF | Length range % | Comments | Reference |
|---|---|---|---|---|
| *D. labrax* | 14 | 2 | juveniles chosen from rearing tanks | Herskin and Steffensen (1998) |
| *L. aurata* | 28.5 | 29.7 | length data from Koutrakis et al. (1994), captured from wild | Marras et al. (2015); Koutrakis et al. (1994) |
| *R. rutilus* | 11.6 | 10.5 | here length % is mean of 7-14 %, captured from wild | Svendsen et al. (2003) |
| *T. mediterraneus* | 29 | | | Zuyev and Belyayev (1970) |
| *Gadus morhua* | | 44.91 | | DeBlois and Rose (1996) |
| *P.phoxinus* | | 39.54 | | Ward and Krause (2001) |
| *L. ramada* | | 24.1 | | Koutrakis et al. (1994) |
| *C. labrosus* | | 40.1 | | Koutrakis et al. (1994) |
| *L. saliens* | | 46 | | Koutrakis et al. (1994) |
| *M. cephalus* | | 16 | | Koutrakis et al. (1994) |

**Table 3**. Summary of energy saving quantities, here tail-beat frequency vs. length ($VO_2$ vs. length data obtained is insufficient for meaningful statistics). Here the 6 quantities in highlighted boxes yield $R^2 = 0.8218$, P-value = 0.7048; Pearson correlation = 0.9065.

While the paucity of data exhibited in Table 3 cannot be conclusively relied upon, it indicates evidence of a relationship between the variation range of fish lengths and the energy saving quantity in terms of tail-beat frequency.

Notwithstanding the apparent relationship between energy saving and variation range as suggested in Table 3, we argue that even taking into account the effects of other schooling behaviors and some imprecision among measures of the energy savings quantities, the variation range hypothesis is not strictly a function of the energy saving quantity, but is also partly a function of school size. So, for very large schools such as reported by Koutrakis (1994) and DeBlois and Rose (1996), we may predict higher variation ranges as described by Eq. 6 and Eq. 7. This is because the permissible power differential between strong and weak fishes increases according to the size of the school due to the fact weaker fish have more time to fall back within the school, yet maintain proximity to energy saving zones of other fishes.

By adapting Eq. 5, we can model the process in which weaker fishes drift to the back of groups and lose contact. Under this model, when group speed is increased until $GCR > 1$ (the inequality of Eq. 2 is satisfied), fishes will begin drifting backwards since they can no longer keep up to the pacesetter. If this condition is satisfied ($GCR > 1$), a fish ("Fish A") will decelerate within the school as long as the pacesetter's speed either is maintained, or is increased. However, if the pacesetter decelerates such that $GCR < 1$, Fish A can remain within school boundaries, albeit having shifted its position farther toward the rear.

Thus by adding a school size parameter, we can modify the variation range hypothesis and the argument that fishes maximal sustainable outputs will, over time, tend to converge within the range of (1-$d$) * 100 (as a percent), where $d$ is the output coefficient of a following fish in a drafting position relative to a stronger fish in a non-drafting position, and 1-$d$ is the energy saving quantity.

As noted, if at any time the pacesetter's speed fall such that a following fish's output relative to the pacesetter's output is $GCR < 1$ before time (s) exceeds T$gap$ (Eq. 5), the following fish will stay within the boundaries of an existing school. The following equation (Eq. 6) describes the opposite condition, the formation of sub-schools:

$$Sf \iff T_{GCR>1} > T_{gap},  \tag{6}$$

where $Sf$ is the formation of sub-schools; $T_{GCR>1}$ is the duration at which a following fish's output is insufficient to sustain the pacesetters speed, even by drafting.

Thus for large schools or groups, we may modify the variation range such that:

$$VR = ESB + \left[ ESB * \left( \frac{T_{gap}}{T_{gt}} \right) \right], \tag{7}$$

where $VR$ is variation range; $ESB$ is the energy saving benefit as a fraction, i.e. $(1 - d)$ in Eq. 2; $T_{gt}$ is the total time required for a fish at $GCR > 1$ (Eq. 2) to drift from the front of the group to the back before decoupling (i.e. $T_{gt}$ corresponds to the total length of the school). For example, given an $ESB$ of 0.25 (25 %) a fish that takes 30 s to decouple relative to a total time of 60 s required for a fish positioned at the front to decelerate through the entire group and decouple, gives $\frac{T_{gap}}{T_{gt}} = 0.50$ and $VR$ is 0.375. Thus, fishes within a school that shift backward relative to school-mates may be 37.5 % weaker than pacesetting fishes yet still remain within the boundaries of the group as long as condition $Sf$ (Eq. 6) is not satisfied. This implies that high speeds set by strong pacesetters must be sustained for considerable durations before weak fishes will decouple from the group; should the pace slow for any period, weaker fishes have an opportunity to shift forward within the group.

Eq. 7 will hold true only for large groups, the critical minimal size of which is unknown, but is likely to correspond to fatigue rates and the time that leading fishes sustain maximal speeds before a relaxation in pace occurs allowing decelerating weaker fish to stabilize their positions within the school. Thus the exact relationship between the variation range and school size is unknown, and Eq. 7 may require the inclusion of additional parameters.

Nonetheless, the literature and data reviewed in this paper indicate that the sum of any increase in the variation range due to school size and the energy saving quantity generally does not exceed a total variation of 50-60% (Table 6). In the case of small schools for which $VR$ (Eq. 7) is unlikely to apply, like those reported by Herskin and Steffensen (1998), where the length range was ~2% for juveniles, it is not clear why the variation range would be substantially smaller than the proportionate energy saving quantity. Nevertheless, it is consistent with the variation range hypothesis that the range be smaller, but not greater than the proportionate energy saving quantity (except for the increases that account for school length), since organisms within this smaller range in terms of strength will be capable of sustaining the speeds or outputs of pacesetters. Such a narrow variation range of ~2% for juveniles implies either, or both, a correspondingly low energy saving quantity and frequent periods of high speed that produce sorting within this narrow range. A low variation range for juveniles may also be related to the egg size variation. The implications of this for the variation range hypothesis bear further study.

The question also arises why, for small schools (or any given biological collective), the body-size variation range might be significantly *higher* than the proportionate energy saving quantity. Indeed, we expect that many animal collectives will contain infants and juveniles whose maximal sustainable outputs are well below the hypothesized variation range. This problem does not weaken the logic of the variation range hypothesis, but rather points to two evolutionary solutions, or a combination thereof: first, some form of parental energetic investment in which infants and juveniles are carried (often, quite literally) by parents or stronger group members during migratory movement until the maximal sustainable outputs of these young, weaker members, increases sufficiently to attain the proportionate lower boundary of the energy saving quantity ((1-d) * 100); secondly, that animal collectives maintain a period of comparatively sedentary or slow moving mobility during which young members grow in strength prior to migration. In this way, parental investment compensates for the difference between the maximal sustainable outputs of the young members and the lower boundary of the hypothesized variation range, while a sedentary or slow moving growth period simply allows the young time to grow to this lower boundary before migration. We explore this compensatory investment further in the case of dolphins, and have discussed previously the sedentary juvenile growth period prior to migration in the case of geese (Menu et al., 2005). In the section that follows we explore this further in the case of eels.

From this we may predict that when groups begin their migration, their size or strength variation ranges may be broader than their energy saving quantity because the group will contain weaker juveniles which begin their journey before they are physiologically capable of remaining coupled to the group. Over the course of their migration, however, the group variation range will tend to converge toward the energy saving quantity as weaker animals are filtered from the group, thus potentially narrowing the adaptive strength or size range of remaining group members, or causing sub-group formation in which sub-groups contain members whose strength or size ranges are narrowed (e.g. Fig.8).

## 9. Eels

In contrast to schools of caudal fish (fin swimmers) which swim in rhomboid or diamond patterns, anguilliform (undulatory) swimmers like eels swim in close contact and parallel to each other, temporarily synchronizing undulations (Burgerhout et al., 2013).

Burgerhout et al. (2013) studied European eels (*Anguilla anguilla*) which are capable of migrating distances of up to 6000 km. In their study, the authors first determined that 0.9 m s$^{-1}$ was approximately the eels' maximum sustainable speed. The authors then increased swim tunnel currents from 0.4 to 0.9 m s$^{-1}$ both for solitary eels and for groups of seven, and found a ~30 % reduction in oxygen consumption by eels in a group compared to those swimming in isolation (the energy saving quantity).

There is a reasonable proportionate correspondence between this ~30% reduction in oxygen consumption and eel body-length ranges. DeLeo and Gatto (1995) reported a curvilinear increase in length with age for both male and female yellow and silver eels, as shown in Fig. 9. In Fig. 9 we also show the approximate body-length ranges in terms of a percent difference for each age (not shown in the original).

DeLeo and Gatto (1995) obtained data from eels as they exited a large (10,000 ha) fishery (which they entered from the ocean in the spring) in autumn after metamorphosing into silver eels. The eel samples were taken from a population well-mixed in both age and length; however, the authors reported that each year eels arrive at the fishery from the ocean as elvers (young eels), and they remain for a varying number of years as yellow eels before metamorphosing into silver eels (usually by 8-9 years old for females, and between 4-7 years for males); then, they return to the ocean for spawning. In natural conditions, elvers remain in freshwater regions for years before they migrate to spawn (Ellerby et al., 2001).

From the data in DeLeo and Gatto (1995) we derived a total body-length range of ~76 % (Fig. 9) among all the eels sampled as they exited the freshwater region. However, the critical migratory ages are 4 to 7 years for males, and 8 to 9 years for females; and the body-length ranges of 44 % to 29 % for females of ages 8 and 9, respectively, as shown in Fig. 9 are considerably closer to the indicated 30 % energy saving quantity (Burgerhout, 2013) than the noted total body-length range (~76%). In principle, the potential is high for group divisions to occur if the proportionate variation range exceeds the indicated 30 % energy saving quantity in the case of such long migrations. Because eels do not feed during this migration period (Vøllestad, 1992), their energetic levels diminish at a relatively constant rate; thus differentials in energy levels are more closely related to eels' body size, as opposed to being a result of differential feeding opportunities that arguably have a significant independent influence on how groups divide.

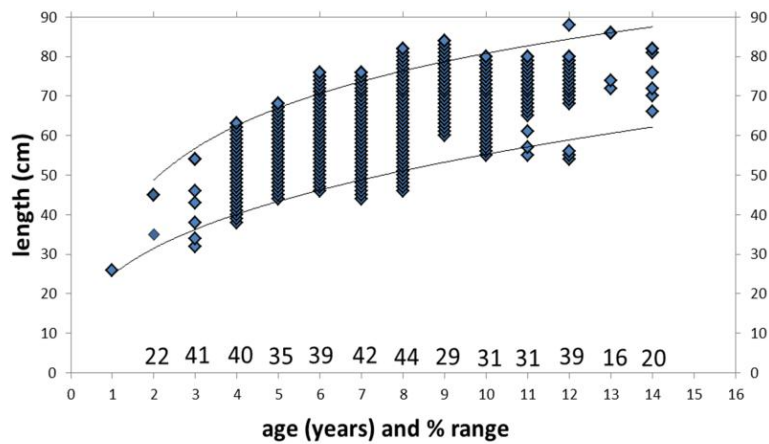

**Figure 9.** Female yellow and silver eel size ranges by age (not distinguished in this reproduction). Data approximated from Fig. 3 in De Leo and Gatto (1995) and adapted to show approximate size ranges. Ranges within age groups show reasonable correspondence to the ~30 % reductions in oxygen consumption (Burgerhout et al., 2013) during group travel. The total size range from the smallest shown to the largest is ~76 %.

Vøllestad (1992) observed that at metamorphosis, which occurred at a mean age of 5.99 years for males and 8.73 years for females, mean male length was 405.6 mm, compared to 623.2 mm for females. Vøllestad (1992) found this 34 % range between males and females difficult to explain particularly without knowledge of eel reproductive behavior, a behavior that has not yet been observed in nature. Vøllestad (1992), however, acknowledged that a male must be of a minimum size to withstand migration stress. We suggest that it is not necessary to observe eel spawning in nature to explain this 34 % difference in size between males and females; indeed this range is expected from the variation range hypothesis and the actual 30% energy saving quantity determined by Burgerhout et al. (2013); this saving enables the weaker males in that range to sustain the speeds of larger, and therefore probably stronger, females.

## 10. Spiny lobster queues

Bill and Herrnkind (1976) reported single-file queuing behavior among spiny lobsters (*Panulirus argus*) which the authors argued is a drag-reduction mechanism. Single-file behavior that involves some recognized or hypothesized energy saving mechanism is also observed, for example, in caribou migrations (Miller et al., 2005), American coots on water (Trenchard, 2013a), ducklings (Fish, 1994), and bicycle pelotons (Trenchard et al. 2014).

Bill and Herrnkind (1976) reported single-file lobster queues of between 2 and 65 individuals migrating at average speeds of 28 cm s$^{-1}$ for periods of over half a day; large queues of 40 or more were observed only over level, unobstructed sand substrate. The authors reported that following lobsters in a group of 19 travelling at 35 cm s$^{-1}$ (the maximum speed observed) enjoyed a 65 % drag reduction (a form of hydrodynamic drafting), relative to lobsters traveling in isolation. The authors noted that drag reduction implies a faster sustainable collective speed, and identified the advantage of drag reduction to weaker lobsters as the facilitation of increased walking speed.

From data in Kanciruk and Herrnkind (1978) of four migration groups adopting single-file formations, we have approximated the carapace length variation ranges for these groups as 45.3 %, 56.7 % , 53.6 % and 52.2 % (from Figure 4, Kanciruk and Herrnkind (1978)). These ranges are in reasonable correspondence with the maximum drag reduction of 65%. This again provides support for the variation range hypothesis, assuming a correlation between lobster size and their maximum sustainable power outputs (MSO) and bearing in mind the caveats regarding the relationship between drag reduction, speed and power output implied by Eq. 3 and Eq. 4.

Bill and Herrnkind (1976) also reported that lobsters will spontaneously trade leading positions, and observed that among groups of 4 to 7 lobsters contained in circular pools, they all shared the leading position. Again this indicates a convective pattern characteristic of energy saving mechanisms in biological systems, including: cyclists (Trenchard et al. 2015), ibis dyadic exchanges (Voelkl, 2015), penguin huddle rotations (Waters 2012; Le Maho, 1977), fish O-turns (Domenici et al., 2002) and rat pup huddle rotations (Alberts, 1978).

Leader position sharing among spiny lobsters is both a good example of protocooperative behavior (Trenchard, 2015), and a simple example of the reasonability of the variation range hypothesis. Under the protocooperative framework, we hypothesize that sharing of the most costly front position can only occur when the speed of the pacesetter leads to the result *GCR* < *d* (Eq. 2) where *d* is the ratio of the maximum sustainable output (MSO) of the follower to the output set by the pacesetter, and 1 - *d* is equal to the energy saving quantity (0.65 for spiny lobsters, and *d* = 0.35).

To illustrate, assuming that a lobster's MSO approximately corresponds to its size (i.e. larger lobsters are stronger and have higher MSOs), we first develop the situation in which following lobsters can sustain the speed of the pacesetter by drafting but are unable to pass the pacesetter (the position locking phase; the stressed regime in Fig. 1). Thus given an energy saving of 0.65 (65%) and a pacesetting lobster travelling at the MSO of the strongest lobster at 35 cm s$^{-1}$, the minimum MSO of a weaker lobster required to maintain the pace of the faster lobster by drafting is 12.25 cm s$^{-1}$ = 35 cm s$^{-1}$ – (35 cm s$^{-1}$ * 0.65), and whose MSO is thus 35% of the strongest lobster (12.25 cm s$^{-1}$ / 35 cm s$^{-1}$).

If 12.25 cm s$^{-1}$ is in fact the MSO of the weaker lobster, then $GCR = 1$ and any subsequent increase in speed by the pacesetter means $GCR > 1$, and the lobsters decouple. When the speed is sustained at 35 cm s$^{-1}$ no lobster can have MSO < 12.25 cm s$^{-1}$ and remain coupled with the group. Moreover, passing can occur only if the MSO of followers exceeds the pace of the leader, such that $GCR < d$. For passing to occur, the leader must either decelerate, or the follower must be stronger: in both cases, $GCR < 1$. This illustrates the decoupling threshold and how it occurs, leading to the hypothesis that the MSO variation range of any biological collective enjoying an energy saving benefit is proportionate to the energy saving benefit because any individual whose MSO is less than a value within this proportionate range will decouple from the group. Thus, in this example, the 65% drag reduction found by Kanciruk and Herrnkind (1978) is proportionate to an MSO range variation of 65%.

In circumstances in which speeds fluctuate between maximum and lower speeds, lobsters with MSO < 12.25 cm s$^{-1}$ may reintegrate after decoupling from the group when the pace decelerates sufficiently; however, this will be highly fatiguing if speeds oscillate and decoupled individuals are continuously forced to catch up to the group ahead. Eventually such individuals will decouple permanently, especially over long migration periods and in relatively small groups, as indicated by Eq. 6 and Eq. 7. If groups decouple permanently such that sub-groups form and reproduce among themselves in isolated geographical locales, the potential for speciation occurs. When this process is repeated over geologic time, the entire diversity of species emerges such that each species exhibits its own unique metabolic output variation range, usually in the form of size variation, corresponding proportionately to the available energy saving quantity.

As discussed, in large schools of fish a greater variation range may exist because of the longer period experienced by weaker fish to drift toward the back of a group (Eq. 7). However, lobster queues are typically less than 40 lobsters long (Bill and Herrnkind, 1978), forming a group length that is probably too small to permit lobsters which are significantly weaker than 35% of the strongest lobsters to remain within the group through an entire migration period.

## 11. Penguin and other huddles

Alberts (1978) was perhaps one of the first to liken collective animal behavior to a "convection current" in describing the rotational behavior of huddling rat pups (*Rattus norvegicus*), or "pup flow". He described the process as one of cooler pups in a huddle burrowing downward into the insulated region of the group, which displaces other pups to peripheries; by contrast, warm pups ascend to the surface, shifting the direction of collective movement, as shown in Fig. 10 (from Glancy et al. 2015).

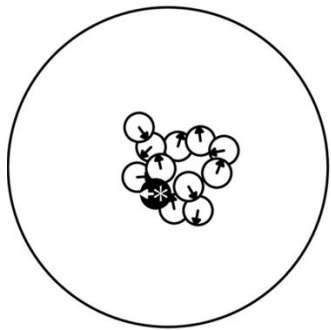

**Figure 10.** Huddling rat pups in a convective process; arrows indicate direction of movement relative to the focal pup, black circle (reproduced from Glancy et al., 2015, *Creative Commons Attribution Licence*).

Glancy et al. (2015) reported an equation for the rate of pup flow (rate of positional exchange) in terms of the time derivative of proportions of exposed surface area. Their equation describes a form of convective flow, yielding a positive value when a pup moves from the centre to the periphery and a negative value when a pup moves from periphery to the centre (Glancy et al., 2015).

Similar descriptions of penguin huddling rotations (Waters et al. 2012; Le Maho, 1977) are consistent with convective behavior. This occurs when huddling penguins shift positions by advancing from windward positions up huddle peripheries, thus sharing the coldest exposed windward positions where heat loss is greatest (Le Maho, 1977). This behavior can be observed in publicly available video (Emperor Penguin, 2016). Stead (2003) applied Le Maho's (1977) description to a computer simulation of this behavior. For a review of huddling behavior and its reductions in metabolic costs see Gilbert et al. (2010), showing metabolic saving from huddling animals of between 6 % and 53 % for 28 different species. In terms of timescale, rotational flow is in the order of hours, longer than other convective motion among animals we have discussed, but vastly shorter than any convective motion that generates phyllotactic arrangements, discussed previously.

A different description of huddling rotations is from Zittebart et al. (2011) who described standing wave motion and "treadmilling" behavior in which penguins join the huddle at the trailing edge and leave it at the leading edge. Similar rotational spiral movement as an element of wave-front motion was simulated by Gerum et al. (2013). Le Maho's (1977) description, Zittebart's (2011) treadmilling description, and Gerum et al.'s (2013) spiral motion description, appear to be similar convective processes in different directions, generated according to different initial conditions that may induce preferential rotational flow in one direction or another.

Trenchard (2013) observed similar collective rotations and sharing of the anterior (lead) position among American s (*Fulica americana*) in broad planar formations, as shown in Fig. 11, and single-file on-water formations. Due to near proximity surface swimming and similar hydrodynamic forces as experienced by mallard ducklings (*Anas platyryhncos*) on water (Fish, 1995), Trenchard (2013a) inferred that American coots on water are likely to experience similar metabolic reductions to those of ducklings on water. Fish (1994) examined on-water duckling formations and found that compared to solitary ducklings, those in single file behind a decoy "mother" experienced 7.8 – 43.5 % reduction in metabolic output, and a maximum of 64 % energy saving for ducklings in larger groups. In high density American coot formations, such as those shown in Fig. 11, in which convective patterns occur, we would expect a similar thermographic profile as that shown in Fig. 12, albeit with smaller temperature differentials than that of penguins.

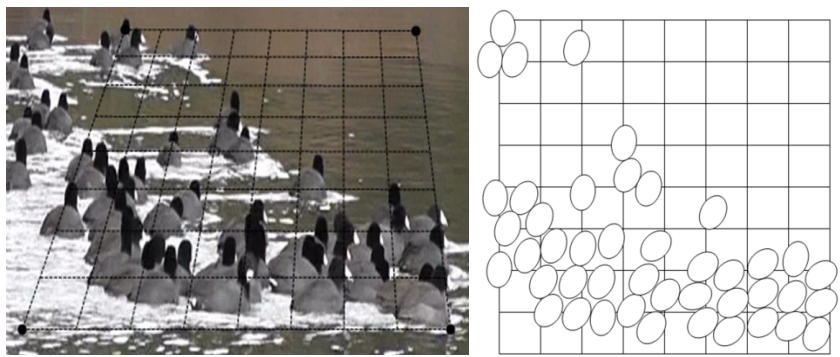

**Figure 11**. American coots in high density formation. Grid approximates overhead perspective. Coots show collective rotational patterns, and given their positional similarity to those of ducks on water which experience substantial energy saving (Fish, 1994), coots are likely to experience similar energy saving by high density on-water formations (reproduced with permission, from Trenchard, 2013a).

Mallards migrate in groups up to thousands of kilometres (Krementz et al., 2011), and coots follow similar migratory patterns to mallards and other waterfowl (Eddleman et al., 1985). Such in flight energy demands may have greater effect on group divisions during longer distance migration than on-water energy demands when birds may be comparatively sedentary or remain within relatively small geographic regions, thus reducing the probability of long-term group separation.

With respect to body size variation ranges, by analyzing data from Owen and Montgomery (1978), we derived ranges of wing measurements for adult and juvenile mallards, male and female: 16.5 % (mean) for the four groups; body-lengths differed by 9.3 % (mean) for the four groups. We have not reviewed studies of the in-flight energy costs and savings of mallards or coots in flight. However, compared with on-water energy saving of ducklings, these wing lengths and body length values are closer to the proportionate in-flight energy saving of 16 % for geese reported by Maeng et al. (2013), and the 14 % energy saving reported by Cutts and Speakman (1984). Nonetheless, further study is required of the combined effects of on-water and in-flight energy saving on variations in size and strength for species that experience both kinds of of energy savings.

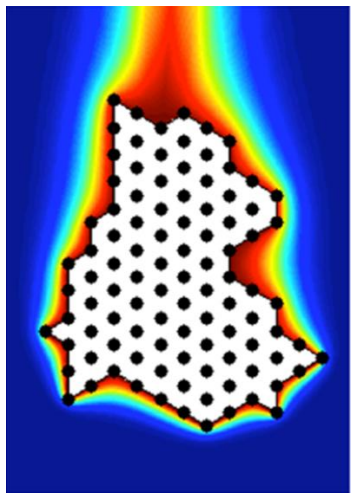

**Figure 12**. Computed temperature distribution around penguin huddle: blue corresponds to cold temperatures; red to warm; individual penguins are black; white indicates polygonal huddle structure (reproduced from Waters et al., 2012, Figure 1, *Creative Commons Attributions* license). Though not stated in Waters et al. (2012), we infer wind direction from the image bottom, with the warmest temperature in the sheltered zone at the top of the image. Note the similarity to Fig. 5(b) in which the temperature distribution here roughly corresponds to what we would expect in Fig. 5(b), except the wind direction is reversed because the cyclists' highest outputs, and corresponding temperatures, are among those at the front in zones of highest drag.

## 12. Dolphins

Weihs (2004) showed that a dolphin newborn (neonate), born at approximately 0.4 the body-length of its mother, exploits Bernoulli suction when it is separated by a few centimeters from the mother's upper body surface and near to its mother's nose; in this position, neonates reduce by 90 % the thrust required to travel at speeds up to 2.4 m s$^{-1}$. For dolphins, Bernoulli suction occurs when the water flow between mother and neonate increases in correspondence to increases in dolphin speed, producing a drop in pressure between mother and neonate, while pressure above the neonate increases (Weihs, 2004). Shoele and Zhu (2015) indicate the optimal suction force position between mother and neonate is about 0.2 of the body-length of the mother. Neonates move to this position immediately following birth and might otherwise "pop like a cork" from the water surface (Weihs, 2004).

In order to induce the Bernoulli effect for their neonates, dolphin mothers must swim at high speed (Weihs, 2004). In this sense mothers "carry" the infant and, when carrying their infant, are capable of sustaining only 76 % of their ordinary speeds when swimming in isolation (Noren, 2008). Noren (2008) posited that it is unlikely that mothers decelerate to accommodate the slower speeds of neonates; rather, mothers decelerate by necessity as a result of the increased drag when carrying the neonate, as evidenced by a 17 % increase in the mother's tail-beat frequency at slower speeds. This is an example of the predicted initial parental investment required to nurture the young up to the lower boundary of the variation range.

Within a few hours of birth, the calf shifts position to the "echelon position", down to a more lateral position within 10 cm of the mother's flank, where energy saving induced by hydrodynamic drafting is about 62 %; at separations of 30 cm is about 25 % and, for older infant calves, energy saving is about 20 % (Weihs, 2004). Noren et al. (2008) reported an average 28% increase in speed for calves in the echelon position compared to swimming in isolation.

The variation range hypothesis predicts that once juveniles reach the lower boundary of the variation range, little or no further parental investment is required. In the case of dolphins, the mother's initial energetic investment to overcome increased drag and to swim faster while the neonate is drawn in close by Bernoulli suction, is followed soon after by the mother's reduced energetic requirements as the calf orientates itself to the echelon position where it can sustain the higher speeds of the mother (Weihs, 2004). This transition from "carrying" to drafting is consistent with the variation range hypothesis, although there may remain some parental investment in terms suckling or protective vigilance. It is noteworthy that the calf is born relatively large – at 0.4 body-length of the mother, and within hours is capable of sustaining the comparatively high speeds of the mother by drafting (Weihs, 2004).

While research has established the energy saving mechanism of drafting and Bernoulli suction between mothers and calves, there appears to be little or no published research that quantifies energy saving mechanisms among adults or among larger dolphin collectives. Connor et al. (2006) suggested that, as an alternative hypothesis to cooperating alliances among bottlenose dolphins staying within 0.5 to 1 m of each other, such close proximity behavior could be an adaptive trait due to drafting. This would be particularly advantageous to pregnant dolphins whose increased frontal surface area and corresponding drag forces cause substantial reductions in speed compared to when dolphins are not pregnant (Noren et al., 2011).

Williams et al. (1992) demonstrated that dolphins and porpoises preferentially seek out bow and stern waves for durations up to hours, exploiting energy saving advantages in these "wave-riding" positions. However, the authors did not present evidence that coupled adult dolphins exploit similar advantages. Similarly, while interactions between dolphins and other cetaceans like right whales have been observed, the research is inconclusive as to whether the purpose of these interactions is to exploit the energy saving mechanism between dolphins and other cetaceans (Wursig and Wursig, 1979).

In a somewhat unique context, Fish et al. (2013) reported that Bowhead whales (*Balaena mysticetus*) display V formation echelons and vortice patterns during coordinated surface feeding behaviors; from these behaviors, Fish et al. (2013) inferred hydrodynamic drafting and energy saving benefits. These behaviors do not appear to have been observed, however, among dolphins.

# 13. Caribou

Miller et al. (2005) observed that caribou (*Rangifer tarandus)* tend to migrate in single-file, with followers stepping in exactly the same spot as those ahead. Fancy and White (1987) suggested this is a behavioral adaptation, although they did not expand upon how this is so, while Miller et al. (2005) suggested this behavior permitted trail marking with scent glands. Mattfeld (1973) (as cited in Boertje (1985)) found that energy requirements for while tailed deer (*Odocoileus virginianus*) walking in snow were 40 % greater than walking on barren ground, and so we may expect Caribou to reduce energy costs by following in the footsteps of those ahead in a similar quantity as for white tailed deer.

Particularly in view that caribou will tend to place their feet where snow has already been compacted by those ahead, the energy saving benefits of single-file walking are strongly indicated. Single-file walking and trail creation are therefore a form of drafting in that leading animals expend greater energy in breaking the snow relative to followers. This is similar to spiny lobsters queue formations, for which Bill and Herrnkind (1976) reported drag reduction benefits. Similarly, Couzin and Krause (2012) modeled pedestrian lane formation behavior in which following pedestrians reduce the energy costs of lateral motion by following in the paths of others.

Fancy and White (1987) reported that the locomotion cost for caribou moving through snow increased exponentially with sinking depth, and that dense snow or crust further increases drag and requires animals to lift their legs higher. The authors noted that for caribou 1 year or older, leg length is a better predictor of net locomotion costs than body weight, and that calves would achieve maximum leg lengths in their second year, despite increasing mass. Fancy and White (1987) noted that the locomotion costs relative to sinking depths are similar for caribou, elk, white-tailed deer, and mule deer.

Duquette (1988) observed that in deep snow conditions, lead caribou appear reluctant to break the trail, and that caribou are often observed to walk routes that have previously been travelled by upwards of thousands of caribou. Presumably, caribou must break trails relatively frequently, although the presence of pre-existing trails distributes the energy saving advantages in a complex fashion that may be difficult to model. While there is some indication that leg length is an adaptation to snow depth (Fancy and White, 1987), and a relationship is indicated between the energy saving of followers relative to trail-breaking caribou, further research is required to identify the appropriate parameters to establish a link between caribou size or leg length, and their path creation and drafting behavior. In terms of aerodynamic drafting between caribou, quantities may be similar to those determined by Spence (2012) for race-horses.

# 14. Turtle hatchlings and semi-fluid dwellers

An emerging area of study involves the motion of animals through semi-fluids like sand or wet soil (Dorgan, 2015). Rusli and Booth (2016) studied the upward digging of Brisbane river turtle hatchlings (*Emydura macquarii signata*) through moist sand. Using respirometers to measure hatchling $CO_2$ production, the authors found that larger clutches produced lower total $CO_2$ when ascending to the sand surface. In a clutch of 4, the slowest turtle took 162.77 hours to emerge from the sand, while the fastest took 112.25 h to emerge. In a clutch of 14, the slowest turtle took 49.94 hours and the fastest took 39.94 h to emerge. The authors reported that a hatchling digging on its own would consume ~2.05 kJ compared to 0.59 kJ in a clutch of 14; a ~71 % energy saving for *E. signata* hatchlings.

Larger groups required less resting time between bouts of digging, which the authors suggested was probably due to lower blood lactate accumulated during digging periods. The authors reported the formation of discreet groups that engaged in synchronous digging activity, which could be triggered by any member of the group. The cue used to trigger the digging activity is unknown, but the authors proposed that it may be related to the fall of blood lactate levels below a critical threshold. The actual energy saving mechanism was not identified, but the movement of multiple turtles through the sand in synchronous motion is consistent with shear and friction force reductions in vibrated sand (Jaeger and Nagel, 1992), which may be the source of the energy saving mechanism.

Rusli and Booth (2016) did not report the size variation among the *E. signata* hatchlings they studied. However, Janzen (1993) showed a correlation between *Chelydra serpentina* hatching size and survival success in moving from nest site to water, supporting the "bigger is better" hypothesis. Similarly, Janzen et al. (2000) made findings for red-eared slider turtles (*Trachemys scripta elegans*), in which smaller hatchlings had lower survival rates than larger ones. In their test procedure, Janzen et al. (2000) did not observe survival rates from the point at which turtles emerged from the sand, but rather the authors released hatchlings systematically on the sand surface. Thus for *T. elegans* we cannot infer a relationship between an energy saving quantity and hatchling sizes or the group structure that may have formed during their vigorous ascent to the sand surface after hatching.

Nevertheless, Janzen et al.'s (2000) data does assist our analysis because it shows *T. elegans* hatchling adaptive size ranges: 34.3 % carapace length range (23-35 mm), and 64.2 % (3.18-8.89 g) mass range. Kolbe and Janzen (2002) reported a somewhat smaller mass range of 54.5 % for 463 snapping turtle (*C. serpentina*) hatchlings upon release. Congdon and van Loben Sels (1990) reported carapace length range of 22.6 % (30.0-38.8 mm) and a similar mass range of 53 % (6-13 g) for Blanding's turtle hatchlings (*Emydoidea blandingii*).

In analyzing these results, we see the carapace length variation for the two turtle species (*T. elegans and E. blandingii*) is not in close correspondence with the proportionate energy saving quantity of ~71 % found by Rusli and Booth (2016) for *E. signata*. However, the mass variations of 64.2 %, 54.5 % and 53 % for *T. scripta*, *C. serpentina*, and *E. blandingii*, respectively, are somewhat closer to the proportionate energy saving quantity for *E. signata*. Although the carapace length data is drawn from species other than *E. signata* – the only species for which the approximate energy saving quantity is known -- there is reasonable consistency among the three mass ranges identified. This suggests that similar mass ranges may be present among *E. signata* and therefore that these ranges reasonably correspond to the approximate energy saving of 71% shown by Rusli and Booth (2016) for *E. signata,* as predicted by the variation range hypothesis.

We reiterate that the relationship between body-length, mass, and the energy saving quantity is uncertain. In our review of the fish schooling literature, for example, the fishes' body-length ranges appeared to be in closer agreement to the energy saving quantity than their masses, which may be reasonable because fishes' body-length correlates to tail beat stride length and swim speed (Weihs, 1973; Krause et al., 2005) better than their body mass does. For turtles, however, their masses may be a better indication of the oxygen consumption capacity in the unique sand-fluid dynamics of vibrating sand (Jaeger and Nagel, 1992). Generally, whatever parameter(s) that best correspond to organisms' maximal sustainable power will be the appropriate one(s) by which to determine a correspondence between the variation range and the proportionate energy saving quantity.

## 15. Bacteria

Bacterial collective dynamics involve complex considerations like dipole forces; and, at small separations certain factors dominate hydrodynamic factors, including noise factors, repulsion forces that reduce adhesion, flagellar and lubrication forces (Drescher et al., 2011). In the specific case of *Bacillus subtilis* high bacterial concentrations are shown to involve counterintuitive fluid behavior, and hydrodynamic feedback between the fluid and the bacteria (Wolgemuth, 2008). Further physical dynamics specific to micro-organisms like bacteria and not observed among larger animals include Brownian motion, wall effects, and unique morphological adaptations like rotors, flagella and spontaneous reverse directional capabilities without inertial forces (Condat et al., 2005; Condat and Sibona, 2002).

### 15.1. High collective bacteria speed compared to speed in isolation

The presence of an energy saving mechanism implies that coupled organisms exploiting such a mechanism will probably travel at higher speeds than individuals moving singly in isolation.

Indeed this is true among a number of systems where there is a well understood energy saving mechanism, such as cyclist pelotons (Olds, 1998), inorganic circulating particles (Grujic and Helleso, 2007), and bull spermatozoa (Woolley, 2009).

Because bacteria have been shown to travel faster as a group than when alone in isolation (Cisneros et al., 2007), we may infer an energy saving mechanism, despite that a clear energy saving mechanism does not appear to be revealed in the literature. Cisneros et al. (2007) reported certain speed-density correlations in *B. subtilis,* and observed that aggregates of *B. subtilis* bacteria move faster than isolated individuals.

When individually isolated, *B. subtilis* tend to move at variable speeds, whereas when forming co-directional phalanxes, all members of the phalanx proceed at the same velocity, more than double the speed of individuals: individuals typically swim between 15-30 $\mu$m s$^{-1}$, whereas phalanxes typically swim ~60 $\mu$m s$^{-1}$ (max 150 $\mu$m s$^{-1}$) (Cisneros et al., 2007). Cisneros et al. (2007) indicated that the mechanism for increased collective speeds is the transverse flows between the body of a follower and the tail of a leader; and such flow speeds increase as organisms approach walls and each other.

Mitchell et al. (1995) reported even greater collective bacteria swimming speeds compared to isolated individuals: marine bacteria inside a micro-swarm sustained a speed of 230 $\mu$m s$^{-1}$, about five times the mean speed of 45 $\mu$m s$^{-1}$ shown for a bacterium outside the micro-swarm. Subsequently, Mitchell et al. (1996) showed that when isolated outside a micro-swarm, individual *Shewanella putrefaciens* bacteria achieved mean fastest speeds of 97 $\mu$m s$^{-1}$, whereas inside a micro-swarm, the mean fastest speeds were 187 $\mu$m s$^{-1}$, approximately twice the individual maximums.

Indeed, Mitchell et al. (1995) identified the unresolved extraordinary efficiencies in energy consumption by which micro-swarming bacteria move in speeds at multiples of the predicted 100 % energetic consumption. These efficiencies imply a set of energy saving mechanisms that are not yet well understood.

## 15.2. Bioconvective behavior

In addition to faster collective bacterial speeds relative to the speeds of isolated bacteria, we have also argued that an energy saving mechanism is likely to generate convective behavior as a collective manifestation of dyadic rotations between low and high energy positions, as discussed earlier.

Although known since 1848 (Plesset and Winet, 1974), Platt (1961) appears to have been the first to identify polygonal structures similar to Rayleigh-Bénard cells in micro-organism aggregations due to external stimuli like gravity, light or chemical sources, but not as a thermal convection process (Plessett and Winet, 1974).

For example, *B. subtilis* form three-dimensional bioconvection patterns (Janosi et al., 1998; Kessler et al., 1994), in which critical aggregate densities and fluid Reynolds numbers (Yanoaka et al, 2009) determine a transition to bioconvection. In this state, bacteria swimming vertically to water surfaces for oxygen do so in lower concentrations than plume regions that descend centrally (Yanoaka et al, 2009). Janosi et al. (1998) did not report the ascent speeds of bacterial concentrations relative to descent speeds, although they suggested that swimming speeds are correlated with bacterial cell concentrations. Wolgemuth (2008) described *B. subtilis* roll-like patterns involving collective density fluctuations that precede turbulence, which are similarly reminiscent of convective behavior.

Although not explicitly identified as convective behavior, Mitchell (1996) provided data indicating periods of relatively high bacterial speeds on the edges of micro-swarms relative to speeds inside swarms. This appears to describe horizontal two-dimensional bioconvection as opposed to vertical three-dimensional convection, similar to the collective uni-directional rotational effects observed in penguin huddles (Fig. 4b) and bicycle peloton echelon formations (Fig. 4a).

### 15.3. Consistent with the variation range hypothesis

A prediction that follows from the variation range hypothesis is that when collective speeds are five times maximal individual speeds, as indicated by Mitchell et al. (1995), some bacteria among the collective will be smaller in size or maximal metabolic capacity than their larger or stronger counterparts by a corresponding factor of five, bearing in mind appropriate adjustments for the unique fluid dynamics of bacteria, as discussed.

These approximate size variations are in fact observed. From data in Trueba and Woldringh (1980), 1253 *B. subtilis* cells ranged in length between ~0.5 μm and ~4.8 μm, although within a narrow diameter range of ~0.75 μm and 0.9 μm, grown with a doubling time of 65 min. Similarly, Trueba and Woldringh, (1980) reported 3000 *E. coli* ranged in length between ~1 μm and 3.3 μm, and diameter ranges between 0.4 μm and 0.72 μm. Donachie and Begg (1989) reported similar size ranges for *E. coli* (range ~1 μm to ~7 μm). It is known, however, that under steady state conditions, bacterial sizes are fairly constant but are more variable in the face of environmental challenges (Chien et al., 2012).

Overall, a direct correspondence between bacterial cell size and magnitude of energy saving oversimplifies the complexities of the physical dynamics that drive bacterial metabolic output, as well as oversimplifying the lifecycles and nutrient availabilities of bacteria and their effects on cell growth and size. Nonetheless, increases in collective speeds relative to the speeds of individuals, in addition to the presence of bioconvective behavior, implies an energy saving mechanism. Further, there is some evidence of a correlation between the energy saving mechanism and the size variation ranges of bacteria, as predicted by the variation range hypothesis. Further analysis of existing research and new research is required to determine the extent of this correlation.

# 16. Spermatozoa

The mechanisms by which sperm aggregate, and the evolutionary advantages of these aggregations, are not well understood (Higginson and Pitnick, 2011). Pizzari and Foster (2008) argued that these agglomeration behaviors are examples of sperm cooperation, altruism, and spite. We propose that sperm aggregation behaviors are driven by their underlying energy saving mechanisms, and these mechanisms likely precede such factors as cooperation, altruism and spite in the evolutionary lineage because they emerge from basic physical forces, drag and fluid principles. This proposition is supported to some extent by the statistical finding that head conjugating (connecting) among diving beetle sperm (*Dytiscidae*) is ancestral (Higginson et al., 2012), and that head conjugation is therefore a basic and primordial manifestation of sperm behavior.

Energy saving mechanisms have been reported for a number of sperm aggregates. Woolley et al. (2009) observed that flagella synchronization magnifies beat frequencies and swimming velocities, similar to that observed for eels (Burgerhout et al., 2013). Moore and Taggart (1995) found that, by conjugating their heads, opossum sperm (*Monodelphis domestica*) improve collective speeds relative to solitary individuals. Moore et al. (2002) found wood mouse sperm (*Apodemus sylvaticus*) link together with apical hooks, forming trains of multiple individuals that travel faster together than alone. For a review sperm aggregation properties, see Pizzari and Foster (2008); Higginson and Pitnick (2011).

Hayashi (1998) reported that fishfly spermatozoa (*Parachauliodes japonicas*) form bundles, and as bundle sizes increase, so do bundle speeds; whereas bundle speeds tend to decrease as the medium viscosity increases. Hayashi (1998) noted that it was impossible to count the number of individual sperm within each bundle; thus it is not possible to compare the maximal speeds of individual fishfly sperm in isolation from their bundle speeds. Nonetheless, the correlation between bundle size and speed indicates that bundle formation involves an energy saving mechanism that permits higher collective speeds relative to individual spermatozoa swimming in isolation.

## 16.1. Determining the energy saving quantity

Woolley et al. (2009) noted that sperm synchronize their motion when their heads are in contact and, when synchronized in this way, sperm swim at higher speeds than when each swim alone. As in the case of other biological collectives discussed in this article, we suggest these higher collective speeds indicate the presence of an underlying energy saving mechanism. Thus, in order to determine the energy saving quantity, using data from Woolley et al. (2009) (their Figure 4), we noted sperm speeds while synchronized, and compared these speeds to sperm swim speeds for individual sperm in isolation.

Thus, for two sperm identified in Figure 4 from Woolley et al. (2009) as sperm A and sperm B, we approximated the following for 22 data points: connected sperm A and B were 39% faster than sperm A in isolation; connected sperm A and B were 25% faster than sperm B in isolation; for the two sets, the connected sperm were on average 32% faster than isolated sperm.

## 16.2. Do sperm size variation ranges correspond to the energy saving quantities?

Sperm velocity and size are positively correlated (Mossman et al., 2009; Gomendio and Roldan, 2008), although it should be noted that there is a trade-off between the number of sperm on course to egg fertilization, and their size (Gomendio and Roldan, 2008); and this may affect the degree of correlation between sperm velocity and size. Nonetheless, the variation in sperm sizes within a given spermatophore (single mass ejaculate), appears to correlate reasonably with their swimming strength. Sperm velocity is usually presented as both curvilinear, which accounts for head yawing motion, and as straight-line velocity (e.g. Woolley et al., 2009).

What then are the sperm size variation ranges? Here we seek evidence of variability among individuals within distinct spermatophores so as to compare that variability with the energy saving quantity, and not variations between spermatophores of different males or between species. Although we did not undertake an exhaustive search of the literature, size measurements of sperm from individual spermatophores were components of a number of papers (e.g. Morrow and Gage, 2001; Hettyey and Roberts, 2006; Miller and Pitnick, 2002; Hayashi, 1998), but of those we reviewed, only Birkhead and Fletcher (1995) provided sufficient information to determine sperm size range (for Zebra finch sperm (*Taeniopygia guttata*)), as shown in Table 4.

| Species | Energy saving (%)* $\mu m\ s^{-1}$ | Size range (%) | Reference |
|---|---|---|---|
| Opossum | 28.0 | | Moore and Taggart (1995) |
| Zebra finch | | 27.1 | Birkhead and Fletcher (1995) |
| Bull | 32.0 | | Woolley et al. (2009) |
| Norway rat (lab) | 12.9 | | Immler et al. (2007) |
| Norway rat (wild) | 25.2 | | Immler et al. (2007) |
| House mouse | -33.0** | | Immler et al. (2007) |
| Wood mouse | 34.1 | | Moore et al. (2002) |

**Table 4.** Energy saving quantities among given animal spermatozoa, and size ranges where given or implied. *Straight line or curvilinear trajectories. **Groups were slower than individuals, but the authors suggested that the groups may still exhibit greater thrusting power than individuals.

The values in Table 4 show that the size variation range for Zebra finch sperm reasonably corresponds to the energy saving quantities of four other species' sperm. Although we do not have an energy saving quantity for Zebra finch sperm, the general consistency among the energy saving quantity for the others provided suggests a similar energy saving quantity for Zebra finch sperm. In turn, the values in Table 4 provide some evidence of a correspondence between sperm size variation ranges and the proportionate energy saving quantity.

### 16.3. Sperm protocooperative behavior

We further propose that sperm aggregation is an example of the position locking phase (high output) of protocooperative behavior (the stressed regime in Fig.1). In this phase, individuals exploit the energy saving to sustain the speeds of pacesetters, but are unable to pass them (see discussion on fish schools) (Trenchard, 2015). For cyclists, this phase emerges at the threshold $GCR = d$ (drafting coefficient) when, above this threshold, collective behavior self-organizes as single-file formations (the stressed regime in Fig.1); below this threshold, collective behavior manifests in higher density and as high frequency and magnitude passing behavior (Trenchard, 2015) (the high-density regime in Fig.1).

Because sperm are driven competitively toward egg fertilization and tend to swim at maximal sustainable speeds (Gomendio and Roldan, 2008), a lower output convective phase in which abundant passing occurs is less likely. However, when swimming near maximum speeds, we may expect sperm to operate at a higher output phase in which passing frequency and magnitude are diminished.

Thus sperm travelling near their maximal sustainable outputs will naturally form aggregates and bundles; thus evolving to link and swim together, locked in position, without passing one another. In this way, the bundling phenomenon may be the sperm equivalent of single-file behavior in cyclists when cycling in a high output, low passing frequency phase, as shown in Fig. 13 (the stressed regime in Fig.1). Furthermore, we suggest that sperm aggregations represent an intermediate evolutionary stage in which bundles begin to act as unified individuals (superorganisms) and develop coupled relationships in which additional energy saving may occur between bundles.

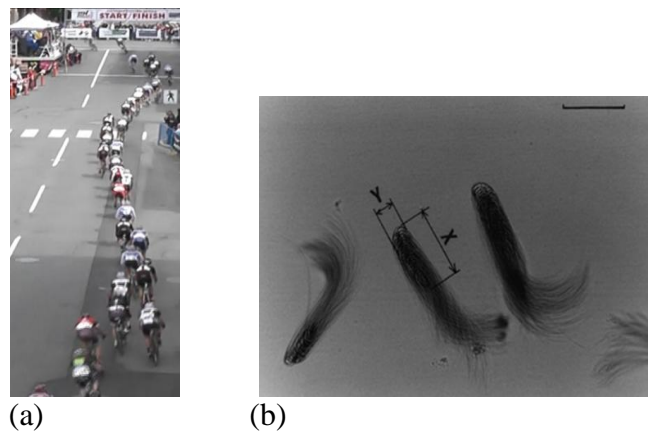

(a)                    (b)

**Figure 13** (a). Cyclists in high-output phase in which passing frequency and magnitude are diminished (image by H. Trenchard); (b). Sperm bundle in apparently similar high-output phase in which passing frequencies and magnitudes are diminished; when sperm sustain speeds of pacesetters by exploiting energy saving mechanisms that arise in close-proximity regions, such bundling may be a natural evolutionary consequence. Bundles may become superorganisms in which higher order group coupling behaviors occur, and drafting between bundles is possible (reproduced with permission, from Hayashi (1998)).

# 17. Krill

Antarctic krill (*Euphausia superba*) aggregate in high density schools up to 20,000 - 30,000 individuals/m$^3$ in which krill orient in parallel (Ritz, 2000); or krill may pack loosely and non-uniformly in swarms up to 1000 m long by 30 m thick (vertical depth), in areas of more than 9000 m$^2$ (Tarling and Thorpe, 2014; Tarling et al. 2009). Krill can swim horizontally at 20 cm s$^{-1}$ (Hamner, 1984) and migrate hundreds of kilometres in a few days (Kanda et al., 1982). Adult Antarctic krill average about 4 cm in length (Patria and Wiese, 2004), and tend to swim at 45° above and below nearest neighbors (O'Brien, 1989).

Ritz (2000) studied energy saving in mysids (*Paramesopodopsis rufa*), which have similarly shrimp-shaped bodies to *E. superba*, between 9 and 13 mm. He found that swarm aggregations up to 50 or 100 mysids consumed on average 7.2 times less oxygen per milligram than non-aggregating individuals. High density schooling can result in reduced oxygen supply (Catton et al., 2011); hence one advantage of the reduced oxygen consumption requirement reported by Ritz (2000) may be to compensate for the reduced oxygen supply associated with high-density schooling.

Ritz (2000) observed that when swimming at high speeds, euphausiids and mysids orient more horizontally in the water, and the aggregations may themselves be a source of updraft that minimizes individuals' efforts to keep from sinking. Unlike pelagic fish, euphausiids and mysids have no buoyant swim bladder and so the energetic costs to sustain swimming to oppose gravity are high, and Ritz (2000) suggested that saving energy by schooling is a necessary selection mechanism.

Ritz (2000) identified the lift-generating vortex rings of neighbors as the energy saving mechanism. Patria and Wiese (2004) studied the field flow of North Atlantic Krill (*Meganyctiphanes norvegica*), a species related to *E. superba*. *M. norvegica* use their 5 pairs of pleopods (swimming legs) to produce a propulsion jet in a circular cross-section, extending 8-10 cm from jet origin, relative to approximate body-lengths of 45 mm, as shown in Fig. 14. The authors observed that a leader and follower tend to synchronize pleopod beat frequencies, and that upward and forward-directed water flow within the vortex represent lift and propulsion which assist following krill to maintain positions relative to the leading shrimp. While this strongly indicates an energy saving benefit for followers, Patria and Wiese (2004) identified both the energetic advantage of these vortices and their utility in short range communication.

Yen et al. (2003) examined similar flow fields of *Euphausia pacifica* and hypothesised the presence of both the energy saving benefit of drafting in the flow induced by neighbors, as well as the signalling effect in revealing the locations of potential mates.

### 17.1. Does the size or output variation range correspond to the energy saving quantity?

If we apply the findings of Ritz (2000) for mysids and hypothesize that the reduction in aerobic requirements, by a factor of 7, corresponds to an energy saving mechanism, we would predict mysids within a seven-fold size range to be present within the same school (assuming a correspondence between size and aerobic capacity). In our review of the literature, however, krill size ranges do not span such a large range (i.e. by a factor of 7) (Table 5). Thus there is no clear correlation between the energy saving quantity reported by Ritz (2000) and krill size ranges; hence these findings do not support the variation range hypothesis.

On the other hand, the lack of correspondence between krill size ranges and the energy saving quantity reported by Ritz (2000) does permit us to question the validity of the energy saving quantity he found. In the first place, Ritz' (2000) reported seven-fold reduction in aerobic requirements, or even similar values, do not appear to be replicated in the literature, despite some evidence for some undefined energy saving indicated by Patria and Wiese (2004) and Yen at al. (2003). It should be noted that Ritz' (2000) values are similar to those found for bacteria. Bacteria, as discussed, however, appear to represent a special case because they involve unique fluid dynamics. In any case, unlike krill, we did observe some correspondence between the energy saving quantity and the size variation range for bacteria.

Consequently, despite the findings of Ritz' (2000) we suggest the question remains unanswered: what is the evidence of an equivalence between ranges of krill power and/or aerobic capacities (expressed in terms of size) and any proportionate energy saving derived from their collective coupling? Conversely, can we look at the size variation ranges, and predict the energy saving quantity? We suggest, based on the circumstantial evidence that we show among other animal collectives for this relationship, it is possible to predict the energy saving quantities from variation ranges, as percentages.

Watkins et al. (1992) studied the size and composition variations among 38 *E. superba* swarms. While the authors did not report the maximum and minimum body-lengths (BL) in each swarm, they did report the mean BL in each swarm and the BL range. The mean BLs per swarm do not allow an accurate computation of the percentage variantion, but we used (% variation = $BL_{range}$ / $BL_{mean}$) to obtain an approximation, returning a mean variation range for 38 swarms of 34.2 %, which will be somewhat lower than the more accurate evaluation (% variation = $BL_{range}$ / $BL_{max}$). This in turn predicts an approximate energy saving benefit of > 34%. This of course implies a correlation between body length and swimming power and/or aerobic capacity, which is reasonable given these correlations among fishes (Weihs, 1973; Krause et al., 2005) but other factors that we do not consider here may also affect power or aerobic capacity.

Nicol (1984) studied size variation among swarms of *M. norvegica.* He reported sizes ranged between 25 mm and 35 mm, a variation of 28.6 %. Nicol also presented two tables summarizing size ranges found in several studies involving *M. norvegica* and other species, which we consolidate to include variation percentages and the findings of Watkins (1992) (Table 5). Thus we predict that the percent variations in Table 5 correspond approximately to the percent energy saving coupling benefit. From these values we therefore predict energy saving quantities among krill to be between ~10% and 58%.

| Species | Size range (mm) | Range (%) (max-min / max; unless stated) | Source |
|---|---|---|---|
| *M. norvegica* | 12-20 | 40 | Patience (1909) |
| *M. norvegica* | 21-29 | 27.5 | McDonald (1927) |
| *M. norvegica* | 28-31 | 9.67 | Fish and Johnson (1937) |
| *M. norvegica* | 11-22 | 40.9 | Einarsson (1945) |
| *M. norvegica* | 25-34 | 26.5 | Cassanova-Soulier (1970) |
| *M. norvegica* | 15-36 | 58.3 | Aitken (1960) |
| *M. norvegica* | 15-28 | 46.4 | Cox (1975) |
| *M. norvegica* | 25-35 | 28.6 | Nicol (1984)* |
| *E. superba* | | 34.2 (range/mean) | Watkins (1994)* |
| *E. pacifica* | 17-19 | 10.5 | Odate (1979) |
| *E. pacifica* | 12-22 | 13.6 | Terazaki (1980) |
| *E. pacifica* | 12-32 | 31.3 | Endo (1981) |
| *E. kronii* | 9-16 | 41.2 | Baker (1970) |
| *T. raschii* | 24-32 | 25.0 | Zelickman (1961) |
| *Thysanoessa longicaudata* | 11-16 | 31.3 | Forsyth and Jones (1986) |

**Table 5.** Consolidation of Tables 1 and 2 from Nicol (1984), with the addition of variation percentages, and the result derived from data in Watkins (1994). We predict that the variation percentages approximately correspond to the percent energy saving obtained from the coupled energy saving mechanism. Except for *, citations are as found in Nicol (1984).

## 17.2. Travelling faster as a group

As discussed, an energy saving mechanism will tend to increase the travel speeds of individuals over isolated individuals. Kawaguchi et al (2010) studied krill behavior in holding tanks, and reported a school of eight krill swam on average ~20 cm s$^{-1}$ compared to ~10 cm s$^{-1}$ for a non-schooling group of nine. The authors did not report the speeds of individuals, however, and did not specify whether the individual speeds achieved were maximal sustainable speeds.

Catton et al. (2011), studied small groups between 3 and 6 krill of two types, *E. pacifica*, and *E. superba* in still-water tanks of diameter 600 mm and height 400 mm. The authors reported slightly *slower* mean speeds for the groups than individuals: means for solitary *E. superba* ~7.7 cm s$^{-1}$; group *E. superba* ~6.9 cm s$^{-1}$. The authors noted that inter-individual variation in speed was large enough that mean values were not significantly different. Further, the authors noted that in their study, typically *E. superba* exited the access pipe and swam in a straight line to the other side of the tank before dropping to the tank bottom, limiting data acquisition. Additionally, the results of Catton et al. (2011) were based on small krill sample sizes in relatively small tanks, and their results do not permit conclusions as to differences between group swim speeds and solitary swim speeds, nor do they provide insight into saving in collective hovering costs which Ritz (2000) demonstrated to be substantial relative to non-swarming *P. Rufa*. In general, the literature appears inconclusive as to whether krill swarms are capable of faster swimming than isolated individuals.

Within schools, individuals in leading positions generally incur a higher energy cost than followers which exploit available energy saving mechanisms, as we have discussed in previous sections. For North Atlantic krill (*M. Norvegica*), a close relative of *E. superba* (Patria and Wiese, 2004), the energetically beneficial vortice zones are approximately 30 – 60° below and 0 - 30° above the krill body horizontal axis and where the propulsive jet flows between this region, as shown in Fig. 14. This suggests that at speeds approaching krill maximal sustainable outputs, following krill will gravitate toward vortice uplifts, generating a phase change in collective behavior analogous to phase changes observed in bicycle pelotons as speeds increase toward collective power output thresholds (Trenchard et al., 2015).

## 17.3. Spatial relationships and collective rotations

In studying spatial relationships between various krill species, O'Brien (1989) observed that at relatively low speeds, *E. superba* showed a strong preference for being positioned in the same plane as neighbours, at 0° elevation along the horizontal body axis, as illustrated in Fig. 14. O'Brien's (1989) study did not include incrementally increased current (swimming) speeds for *E. superba*, although he did induce higher swimming speeds for *Tasmanomysis oculata*, *Nictyphanes australis*, and *Teganymysis*, and observed that as speeds increased there was a tendency for each of these to shift preferred positions to ~30 – 60° above and below neighbors, while *Teganymysis* also showed increased preference toward even axis positioning. Such a shift in position when swimming speeds rise, is similarly consistent with animals moving to optimal energy saving zones, preceding a phase change in collective dynamics.

## 17.4. Direction of collective flow

Although there appears to be no existing literature that describes krill formations specifically as convective processes, there are results within the literature that are consistent with convective processes. Using a dye plume to trace mysid directional movement in a holding tank, Ritz (2000) observed a collective clockwise rotation among the currents generated by the krill. Tarling and Thorpe (2014) modeled clockwise and counter-clockwise krill swarm rotations, relative to the background ocean water flow. These rotations produce approximate convective sinusoidal trajectories, an example of which is shown in Fig. 3. As discussed, we suggest these are signatures of energy saving mechanisms, representing large-scale or multi-agent versions of dyadic oscillations, similar to those observed in Northern Bald ibis flocks, that exhibit dyadic oscillations (Voelkl, 2015), or to "two-up" positional alternations that are well known in the sport of cycling.

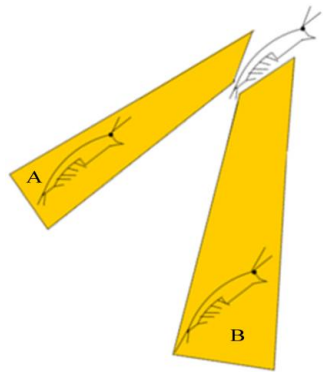

**Figure 14.** Krill energy saving zones. Adapted from images from Patria and Wiese (2004), colored cones (color online) represent approximate zones of upwash vortices for the vertically ascending leader. The unmarked region between the cones is a jet flow, where following costs are increased. Vertically directed krill A is likely to accelerate in the vortice uplift over the top of the leading krill, while krill B accelerates beneath the leading krill. Collective rotation may be predicted in the direction of body orientation, consistent with findings of Ritz (2000) for mysids.

Tarling and Thorpe (2014) discussed the effects of hydrodynamic drag on krill speeds moving with background water flow and moving against background flow, and referred to the standard drag Eq. 3. The authors observed that the when working against the background flow, krill speeds decreased with increasing background flow velocity. The authors suggested this means the krill are acting in some way either to reduce the drag force, or to work harder to maintain speeds.

In terms of increased energetic costs to krill at school leading edges where drag is higher relative to zones of reduced drag, there is some evidence of this in diel migrations. Diel migrations are 24-hour cycles in which krill descend at dawn to depths in the hundreds of meters, and then ascend to the water surface at dusk, or sometimes vice versa (Tarling et al., 2001). Tarling et al. (2001) reported that northern krill (*M. norvegica*) and pteropods (*Cavolina inflexa*) at the leading edge of a swarm during an ascending phase tend to travel more slowly relative to those at median depths. The authors suggested this may be a behavioral response that enables cohesive swarms.

Wiebe et al. (1992) also reported this effect for euphausiid swarms that consisted of a six euphausiid species. Heywood (1996) reported this effect for zooplankton, indicating a kind of compression effect among both the upward and downward direction swarms in which leading edge animals move more slowly relative to followers, thus maintaining higher group cohesion.

We suggest that this behavior is in fact what we would expect in the presence of an energy saving mechanism. For example, even for dead krill descending collectively through water at 2.5 cm s$^{-1}$ (Kils, 1985) we would expect a sedimentation effect in which a "drafting, tumbling, kissing" process occurs repeatedly, as is demonstrated among inert spheres in fluid (Wang and Guo, 2014). However, the effect is exaggerated with increased speeds and, with an energy saving mechanism, we would expect a compression effect to occur in both directions, particularly when leaders are working nearer to maximum while followers exert relatively lower effort to sustain the speeds of leaders; i.e. leaders are effectively pushed backwards relative to followers, generating increased compression.

## 18. Summary

There are various mechanisms in biological systems by which the interactions of coupled organisms facilitate energy saving for at least one member of the coupled system. We have identified a few such mechanisms. Moreover, the occurrence of drafting in non-biological systems indicates the process is a basic physical one that represents a basin of attraction in the energetic landscape, consistent with thermodynamic principles and a tendency for systems to maximize entropy. All of these mechanisms provide the beneficiaries with varying magnitudes of energy saving, in terms of percentages, and the literature indicates these quantities range from ~7 % to 70 %, with some exceptions. Broadly the size or mass variation ranges follow in remarkable accordance with values as predicted by the variation range hypothesis. See Table 6 for a summary.

We have identified certain fundamental dynamical processes that flow from the presence of energy saving mechanisms. One of these processes involves convective effects, characterized by high frequency and magnitude passing among organisms at sufficiently low collective outputs (Trenchard, 2015). Convective behavior self-organizes over short timescales in a variety of biological systems, including bicycle pelotons, bird flocks, fish schools, penguin huddles, krill, and bacteria. Over long time scales, convective behavior occurs in phyllotactic formations. The convective processes are a dynamical phase, and a property that may represent a class of systems.

When speeds increase above a critical threshold but below a decoupling threshold, the frequency and magnitude of passing diminishes, and coupled organisms lock their positions and sustain the speeds of those ahead by exploiting the energy saving mechanism while being unable to pass or share the most costly front positions. This is observed in pelotons, for example, when cyclists are proceeding at near maximal outputs. We propose that sperm agglomerations are examples of this process in nature, since sperm competitively advance at near maximal speeds, thereby locking in relative positions through the equalizing effect of an energy saving mechanism.

We propose the position locking resulting from the high output phase in the presence of the energy saving mechanism precedes sperm agglomerative behavior in its evolutionary lineage. This high output phase resides in the stressed region of the thermodynamic branch, as shown in Fig. 1. This phase is also indicative of a rigid state, as described by Csermely (2015), in which resources are strained due to maximal energy or resource consumption. By contrast, the convective phase exhibits higher degree plasticity (Csermely, 2015) due to its greater abundance of accessible energetic resources and fluidity.

In addition, we have adduced evidence for the variation range hypothesis, perhaps first proposed by Makoto in 1970 in the specific case of fishes. Trenchard (2015) demonstrated by computer simulation that the variation range of cyclists in a peloton (group of cyclists) tends to converge on the equivalent proportionate energy saving when cyclists are driven to their near maximal speeds. At these high speeds, decoupling occurs between cyclists, causing the formation of sub-groups that comprise weaker members who can sustain the speeds of stronger members by exploiting the energy saving mechanism of drafting.

We propose this is a universal sorting principle that applies when collectives are driven to near maximal speeds. The principle is independent of the causes of these near maximal speeds, whether such speeds are, for example, a result of being chased by predators, a foraging effect, or a scaling effect in which collective strength diminishes from fatigue over long range migrations. We suggest this group sorting process is an evolutionary principle that can lead to speciation; and, when repeated continuously over geologic time, the entire diversity of species emerges in which each species exhibits its own unique maximal strength variation range, usually in the form of size or mass, that corresponds to the available energy saving quantity.

We have provided some evidence of a correlation between the size ranges of a variety of species and the proportionate energy saving, generated when organisms are sufficiently coupled. This assumes that size, in terms of body-length or mass, correlates with strength, as the evidence bears out. We have modelled the circumstance in which the size of the group modulates the variation range, because weaker individuals may decelerate relative to others and remain within the system boundaries and within proximity of energy saving zones. Thus if the mean group speed falls, weaker individuals have opportunities to remain within the group. It is not clear how much the size of the system (flock, school, etc.) affects the variation range.

One implication of the variation range hypothesis is that when nascent offspring are too weak to fall within the variation range, mothers and/or conspecifics must compensate for the offspring's weakness by increasing their own energy investment sufficient to ensure offspring survive *until* the offspring's physiological capacity reaches the lower boundary of the variation range. Similarly, in other circumstances, weak young must grow in relatively sedentary conditions for some extended period so as to grow to the minimum boundary of the variation range in order to undertake high cost migrations. Once offspring grow to achieve this minimum physiological capacity, they capably sustain the pace set by stronger members of the collective without further conspecific investment. This has implications for increased understanding the timing of migratory patterns.

## Acknowledgments

The authors wish to thank Demian Seale for his editing assistance; Shaun Killen, Paolo Domenici, Jon Svendsen, for their input on aspects of the section on fishes; and three anonymous referees for their helpful comments.

| System | Energy saving mechanism | Indicated energy saving (% difference except where given as multiple) | Length range; Mass where noted (% difference except where given as multiple) | Pacesetter equivalence | Reference |
|---|---|---|---|---|---|
| **Non-biological** | | | | | |
| Particles in optical vortice | hydrodynamic drafting | 15 | | leading particle | Grujic and Helleso (2007) |
| Particles in chain | hydrodynamic drafting | 50 | | leading particle | Šiler, et al. (2012) |
| **Biological** | | | | | |
| Bacteria *Bacillus subtilius* | transverse vortices | five-fold | up to ten-fold | front position | Cisneros et al (2007); Trueba and Woldringh (1980) |
| Cyclists | aerodynamic drafting | 39 | 36 | front position | McCole et al. (1990) |
| Caribou | trail breaking | 40 | | front position | Mattfeld (1973) |
| Dolphins | Bernoulli suction; hydrodynamic drafting | up to 90 (neonates) 20-28 (older infants) | | mother | Noren et al. (2008) Weihs (2003) |
| Duckling on water | hydrodynamic drafting | 7.8–62.8 | 16.5 (wing length, juveniles and adults) 9.3 (body-length, juveniles and adults) | decoy (mother) | Fish (1994); Owen and Montgomery (1978) |
| Eels (European) | undulatory synchronization | 30 | 34; 29-44 | parallel motion | Burgerhout et al. (2013); Vøllestad (1992) |
| Fish (general) | hydrodynamic drafting; | | ~30 | front position | Pitcher and Parrish (1993) |

| | Karman gait | | | | |
|---|---|---|---|---|---|
| Fish (horse mackerel) | hydrodynamic drafting; Karman gait | 15-29 | | front position | Zuyev and Belyayev (1970) |
| Fish (Grey mullet) | hydrodynamic drafting; Karman gait | 28.5 8.8-19.4 | 29.7 | front position | Marras et al. (2015) Koutrakis et al. (1994) [see table 1 for further quantities] |
| Fish (Roach) | hydrodynamic drafting; Karman gait | 7.3 11.9 11.6 | 7-14 17-41 (mass) | front position | Svendsen et al. (2003) |
| Fish (Sea bass) | hydrodynamic drafting; Karman gait | 9-14 9-23 | | front position | Herskin and Steffensen (1998) |
| Fish (Euro. Minnow) | hydrodynamic drafting; Karman gait | | 39.5 | front position | Ward and Krause (2001) |
| Krill (*E. Superba*) | hydrodynamic uplift | up to seven-fold | 34 | front position | Ritz (2000); Watkins et al. (1992) (see Table 5 for further quantities) |
| Locusts | wing-beat coupling | 16 | | front position | Camhi et al. (1995) |
| Northern bald ibis | vortice upwash | | | front position | Voelkl (2015) |
| Geese (Canada) | vortice upwash | 36 | | front position | Hainsworth (1987) |
| Geese (Canada) simulation | vortice upwash | 16 | | front position | Maeng et al. (2013) |
| Geese (Pink-footed) | vortice upwash | 14 | | front position | Cutts and Speakman (1984) |
| Pelicans (Brown) | vortice upwash (ground effect) | 49 | | front position | Hainsworth (1988) |
| Pelicans (white) | vortice upwash | 11.4-14 | | front position | Weimerskirch et al. (2001) |
| Penguins (Emperor) | huddling | 51 | | birds peripherally | Gilbert et al. (2008) (see Gilbert et al. (2010) for review of huddling |

| | | | | exposed to wind | systems) |
|---|---|---|---|---|---|
| Spermatozoa (Bull) | flagellar synchronization | 32 | | front position | Woolley et al. (2009) |
| Spermatozoa (Zebra finch) | | | 27 | front position | Birkhead and Fletcher (1995) [see Table 4 for further species] |
| Spiny lobsters | queue formation | 65 | 45 54 57 54 52 | front position | Bill and Herrnkind (1976); Kanciruk and Herrnkind (1978) |
| Turtle hatchlings | granular fluid drafting | 71 | 53 55 64 (mass) | front position | Rusli and Booth (2016); Kolbe and Janzen (2002) |

**Table 6**. Summary of approximate energy saving quantities and approximate variation ranges in coupled systems.